\documentclass[aps,amsmath,amssymb]{revtex4}

\usepackage{subfigure,graphicx}

\begin{document}

\title{Dynamics of many-particle fragmentation in a Cellular Automaton model}

\date{\today}

\author{A.\ Lejeune\thanks { Electronic address : a.lejeune@ulg.ac.be}}
\address{Universit\'e de Li\`ege,Institut de Physique, B5 Sart-Tilman,
B-4000 Li\`ege 1, Belgium}
\author{J.\ Perdang\thanks { Electronic address :
jperdang@solar.stanford.edu} } \address{Universit\'e de Li\`ege,
Astrophysique, B5c Sart-Tilman, B-4000 Li\`ege 1, Belgium}
  \address{and New Hall College,Huntingdon Road,Cambridge, CB3 ODF, United
Kingdom}
\author{J.\ Richert\thanks{ Electronic address : richert@lptl.u-strasbg.fr} }
\address{Universit\'e Louis Pasteur, Laboratoire de Physique Th\'eorique,
3, Rue
 de l'Universit\'e,}
\address{F-67084 Strasbourg Cedex, France}

\begin{abstract}

A 3D Cellular Automaton model developed by the authors to deal with the
dynamics of N--body interactions has been adapted to investigate the
head--on collision of two identical bound clusters of particles, and the
ensuing process of fragmentation. The range of impact energies
is chosen low enough, to secure
that a compound bound cluster can be formed.  The model is devised to
simulate the laboratory set--up of fragmentation experiments as
monitored by 4$\pi$ detectors.  The particles interact via a
Lennard--Jones potential.
At low impact energies the numerical
experiments following the dynamics of the individual particles
indicate a phase of energy sharing among all the particles of the
compound cluster.  Fragments of all sizes are then found to evaporate from the
latter cluster. The cluster sizes, measured in our set--up by simulated
4$\pi$ detectors, conform to a power law of exponent $\approx$ 2.6.

In an attempt to duplicate
the laboratory caloric curves related in particular to nuclear fragmentation
processes, we introduce several temperature parameters (kinetic
temperature of nucleons, kinetic temperature of fragments, reaction
equilibrium temperatures).  Theoretical caloric curves are then
constructed for those temperature parameters we regard as physically
most relevant.  Our results show that different temperature
definitions generate different curve patterns, indicating that the
fragmentation system remains far from thermodynamic equilibrium.
The pattern of the laboratory caloric curve
for Au--Au collision experiments as derived from a recent analysis [NuPECC
report Dec.  1997] is reproduced qualitatively by our reaction
temperatures.
\end{abstract}

\pacs{ 05.45.-a, 25.70.z, 87.18.Bb}

\maketitle

\section{Introduction}

The currently favoured theoretical approach of investigating many-particle
fragmentation consists in applying statistical mechanics to the excited
compound cluster. The procedure relies on the assumption
that the collection of particles, or compound cluster, originating from the two
colliding clusters reaches a state of thermodynamic equilibrium.  The
properties of this compound cluster can then be described
thermodynamically, in terms of a small number of global parameters,
namely the total number of particles $A$, possibly the total charge
number $Z$, the total available energy $E$ (regarded as measured in
the centre of mass system), the total volume $V$ occupied by the
particles, and possibly the area of the surface $\Sigma$ enclosing this volume
(\cite{RI01} for a recent review of the theoretical approaches in the
nuclear field; \cite{CH97,FA97,GO01,SC01,SC97,KU99,BR89} for experiments
and results on molecules and clusters of ions).
In this Paper we shall describe a less restrictive theoretical framework
applicable to a multiple fragmentation process of arbitrary nature (nuclear,
atomic, or molecular).  For our specific illustrations the relevant
orders of magnitude of the global parameters have been chosen to
capture the particular nuclear case.

A few comments on the related
laboratory experiments are therefore in order.  The
empirical results are collected essentially through the following
experimental procedure.  A target nucleus ($A_T,Z_T$) is bombarded
with a high--energy beam of incident nuclei ($A_I,Z_I$) generated by
an accelerator.  A series of detector set--ups (such as ALADIN,
CHIMERA, EOS, FOPI, INDRA, LASSA or MINIBALL), ideally uniformly
distributed over a sphere centred on the target nucleus and thereby
securing a $4 \pi$ coverage, identify the charge and mass of the
collisional fragments as well as their kinetic energies.

Experiments of this nature were spearheaded in the 80s
\cite{MI82,PA84,PA85}. In \cite{MI82} the target was
a Kr or Xe ion; the accelerated incident particles consisted in a beam of
protons of energy in the range 80--350 MeV. The detectors isolated
nuclear fragments in the $A$--range 12 $\le A \le$ 31.
The yield $Y(A,Z)$ was found to be
consistent with an expression of the form $A^{-\tau}$ times a
Boltzmann factor \cite{MI82} depending on a temperature parameter.
The exponents evaluated experimentally were $\tau$ = 2.64 (Xe) and 2.65 (Kr).
Qualitatively the outcome of these experiments was compared with
the thermodynamic transition from a liquid phase (original target
heated by the infalling beam of protons) to a state of formation of
droplets of all sizes (fragment distribution measured by the
detectors) taking place at a critical--point temperature $T_c$.  In
fact, for the latter transition a power--law in the sizes of the
droplets is known to hold, of exponent $\tau$ = 2.33.  This result is
derived in the context of the mean field theory \cite{JA83}.  Similar
fragmentation experiments were described in \cite{PA84},\cite{PA85},
in which the target was an Ag, Kr, Xe or U ion; the incident particles
were protons or carbon ions.  Again fragments were detected (here in
the $Z$--range 3 $\le Z \le$ 22), which obeyed an approximate power
law.

The seemingly canonical interpretation of these earlier
as well as of the more recent
intermediate--mass ion--ion collision experiments
(for instance the  $^{197}$Au--$^{197}$Au fragmentation reported on
in \cite{PO95},\cite{NU97})
regards the fragmentation process as the formal analogue
of a liquid--gas phase transition occurring at a transition temperature
$T_{c}$. At lower temperatures, $T <  T_{c}$,
the nuclear system is a mixture of individual nucleons and fragments (liquid
phase); at
higher temperatures, $T >  T_{c}$, all fragments
dissolve into a gas of nucleons (gas phase).

This interpretation of an observed fragmentation process (of any nature) as a
thermodynamic phase transition may be helpful in providing a qualitative
picture for the outcome of the collision experiments.  However, it should be
clear that a many--body fragmentation process induced by violent
collision cannot be regarded as a proper thermodynamic equilibrium
phenomenon.  Prior to the collision, both the target cluster, and the
incident cluster, are in a stable internal equilibrium state.  The two
clusters are characterised globally by two sets of (extensive)
thermodynamic variables $(A_q, Z_q, E_q, V_q, {\Sigma}_q)$, $q$ =
$T,I$.  The energies $E_{q}$ are measured in the centre--of--mass (CM)
system of ions $T$ and $I$ respectively.  The stability of the
equilibrium of each cluster requires a zero surface pressure.  At the
moment of the incipient collision the $T$ and $I$ clusters merge into
a single cluster $C$, the compound cluster $C$ (counterpart of the
compound nucleus $C$ in the sense of Bohr and Wheeler \cite{BO39}).
The extensive thermodynamic variables are then essentially the sums of
the extensive variables of the components $T$ and $I$, $A_C = A_T +
A_I$, etc, except for the difference in the evaluation of the energy.
The energy of $C$, $E_{C}$, is to be measured in the CM frame of the
compound system, so that we have $E_C = E_T + E_I + E_{imp}$; the
extra energy component, $E_{imp}$, is the impact energy (global
kinetic energy of the $T$ and $I$ in CM of $C$).

Since the impact energy is a free parameter,
the surface pressure of $C$, the initial merger of $T$ and $I$,
will not vanish in general, and hence  the compound cluster
will not be in a state of thermal equilibrium.  The surplus energy is
expected to create a pressure increase (positive surface
pressure), which will force $C$ globally to expand.  This expansion may
consist in a release of clusters of particles which ultimately become
independent fragments (as observed in the experiments), leading in
turn to a contraction of a central core.  Such a phenomenon of global
expansion made up of an expanding outer envelope (together with a contracting
inner core) is familiar in gravitationally bound systems (spherical
clusters of stars; cf the classic analysis by Lynden--Bell and Wood
\cite{LY68}).  In the general fragmentation problem, the central
remainder $C'$ is a new compound cluster, with fewer particles, which
keeps evaporating fragments, thereby transforming again into a smaller
central cluster $C''$, as long as it has not reached a final
equilibrium state.  These comments make it manifest that a priori the
distribution of fragments as measured by the laboratory detectors is
not directly related to a statistical equilibrium state of a central
cluster $C$; it is rather the system's non--equilibrium state that
causes the evaporation of the fragments.  However, provided that the
central cluster evolves slowly enough, the component particles may
have time to reach an approximate statistical equilibrium.  The latter
alternative is envisaged in gravitational systems as well.  The
distribution of the clusters that evaporate at a given time $t$ is
then expected to correspond to a chemical reaction equilibrium, or a
dissociation equilibrium, at the temperature characterising the
central cluster at time $t$.  On the other hand, it is clear that this
situation can hold only if the impact energy remains low enough.  At
higher impact energies the target will be essentially transparent to
the particles of the incident cluster.

The simplest statistical models by--pass the conceptual question
of whether or not a statistical equilibrium holds in the
fragmentation process. In these models
the system of particles is enclosed in an energetically insulated box of
finite volume $V_{B}$; the collection of interacting particles is then fully
specified by 3 independent thermodynamic parameters, $V_{B}$, $A_C$,
and $E_C$ (total energy) which are given at the outset.  Such a system
is necessarily due to relax eventually towards a thermodynamic
equilibrium.
In the specific nuclear context, detailed classical molecular dynamics (CMD)
calculations have indeed demonstrated directly the existence of a
statistical equilibrium state, from which the thermodynamic properties
of the system can be recovered \cite{AI88}, \cite{FI86}.  Moreover,
the distribution of the fragments with size can be evaluated in the
thermal equilibrium state.  Similar thermodynamic equilibrium results
have also been derived in the framework of cellular automaton (CA)
numerical experiments (in 2D \cite{AL99} and 3D \cite{AL00}).
However, the connection between these, as well as other theoretical
models so far proposed \cite{RI01} on the one hand, and
the real, a priori far from equilibrium laboratory experiments on the
other hand, remains entirely unclear.

In this Paper we pursue the goal of setting up a theoretical framework
capable of closely simulating the arrangement of the actual laboratory
experiments.  To this end (1) we follow in the first place the
dynamics of the collision (as in the case of CMD calculations) without
relying on an assumption of a thermodynamic equilibrium.  (2) Next we
analyse the collection of fragments which have arrived at a certain
distance $d_D$ from the collision site where they can be regarded as
independent.  The fragment distribution is estimated at that
particular level.  Our procedure of evaluation of the theoretical
distribution thus contrasts with the conventional statistical methods
which have been applied by previous authors, including ourselves
\cite{FI86,AL99,AL00}.  It conforms instead essentially to the
laboratory readings as given by an array of detectors located at a
distance $d_{D}$ from the collision site.  (3) We introduce and
compute explicitly a variety of formal temperature parameters, namely
kinetic temperatures associated with the gas of particles and the gas
of fragments, and reaction temperatures related to the distribution of
the fragments.  If the system were in a state of genuine thermodynamic
equilibrium, in particular during the initial stage of the collision
when the two colliding clusters merge into a single compound cluster,
then a true thermodynamic temperature would exist.  Under this
condition the formal temperature parameters would all be equal to the
true temperature.  Provided only that there are experimental
procedures replicating the conditions of the theoretical definitions
of these formal temperature parameters, the latter continue to provide
a useful global characterisation of the system, whether or not a
thermodynamic equilibrium state is realised: An acceptable theory must
then be capable of duplicating the experimentally available values of
these formal parameters.

The question of the temperatures is analysed in greater detail in
Section 5.  The dynamical calculations (Section 4) are carried out in
the context of the Cellular Automaton (CA) model developed by the
authors and discussed in \cite{AL99} and its 3D extension \cite{AL00}.
A major difference between the numerical experiments of \cite{AL00}
and the experiments of the present work resides in the fact that the
system of component particles was confined to a finite box, while in
principle the system of particles in our present analysis evolves in a
virtually infinite lattice space.

The detailed CA experiments we report on refer to a nuclear
fragmentation involving collisions between identical ions (cf the
$^{197}$Au--$^{197}$Au collision of ALADIN, \cite{PO95}).  We treat these
collisions as being head--on.  In a future work we plan to extend our
model to deal with collisions of non--zero impact parameter.

\section{ The CA Model}

The CA framework adapted to the simulation
of the classical dynamics of interacting particles is
discussed in \cite{AL99}, for the specialised nuclear
context; the particles are nucleons, and we consider low enough energies.  We
indicate here only the particular features of this model when applied
to the collisional dynamics we are concerned with,
and recall the typical orders of magnitude of the model parameters.

\subsection{ Geometry of Lattice Space. Particle Kinematics}

Our CA universe $U$ is a cubic lattice of $L^{3}$ cells,
of toroidal topology (periodic boundaries), with a typical size $L$ = 127.
A lattice cell is identified by a position vector ${\bf r}$ of integer
Cartesian coordinates ${\bf r}$ = $(x,y,z)$, with $x,y,z$ taking the
values $-(L-1)/2$, $- (L-1)/2 + 1$, $\ldots$, -1, 0, 1, $\ldots$,
$(L-1)/2 - 1$, $(L-1)/2$.  An individual cubic cell has a side
${\lambda}$ chosen of the order of magnitude of the range of the
nuclear forces ($\approx$ 2 fm).  The timestep ${\Delta \tau}$ is of
the order of the collision time of nucleons in a bound nucleus
($\approx$ 10$^{-23}$ s).  Time intervals are measured by an integer
$t$ (number of timesteps counted from the beginning of the
experiment).

The real nucleon is simulated by a particle of mass $m$; this
particle is either at rest (symbolized by the zero vector ${\bf e}_0$);
or in a state of motion with a single absolute value of the
velocity
\begin{equation}
v \ \ =  \ \ \frac{\lambda}{\Delta \tau}.
\label{eq:1}
\end{equation}
Accordingly, in our framework the allowed states of motion
of a particle are
{\bf v} = $\pm v {\bf  e}_{i}$, where
${\bf e}_i$  represents the unit vector
along the lattice axis $i$ = $x,y,z$.
A CA particle  $\alpha$ then exists
in one among 7 possible dynamic states
${\bf v}_{\alpha}$ = 0,
$ v {\bf  e}_{x}$,  $ - v {\bf  e}_{x}$,
$ v {\bf  e}_{y}$,  $ - v {\bf  e}_{y}$, or
$ v {\bf  e}_{z}$,  $ - v {\bf  e}_{z}$;
the momentum of this particle is denoted by ${\bf p}_{\alpha}$ (= $m {\bf
v}_{\alpha}$).  As in \cite{AL99} we do not take account of the charge
of the nucleons.

Our CA particles obey an exclusion principle, in the sense
that a cell is not allowed to contain more than one particle
in the same state of motion. Accordingly, the maximum density
of our nuclear matter is 7 particles per cell.

\subsection{Particle Interactions and Dynamics}

To simulate the interactions of a given nucleon $\alpha$
with the rest of the nucleons of our system we choose
an `interaction neighbourhood' of cell ${\bf r}$ containing
nucleon $\alpha$,
$N_{\rm int}({\bf r})$. This neighbourhood is the collection of cells
made up of the central cell ${\bf r}$,
the $\varphi$=6 cells  which have common faces ($\phi$),
the $\epsilon$=12 cells which have common edges (e),
and the $\nu$=8 cells which have common vertices (v)
with the `central' cell ${\bf r}$.
Nucleon $\alpha$ in cell ${\bf r}$ interacts
with any nucleon in a cell  ${\bf r'}$
if and only if $\bf r'$ $\in$  $N_{\rm int}({\bf r})$.

The pair interaction energy between a particle in cell $\bf r$
and a particle in cell $\bf r'$,
$V_{\rm pair}({\bf r},{\bf r'})$, is represented by a step potential:
$  V_{\rm pair}({\bf r},{\bf r'})$ = $V_{\phi}$
if both cells have a
common face;
$  V_{\rm pair}({\bf r},{\bf r'})$ = $V_{e}$, and
$  V_{\rm pair}({\bf r},{\bf r'})$ = $V_{v}$,
if they have a
common edge, or a common vertex respectively.
For a pair of particles in the same cell {\bf r} we adopt a
potential of the form
$V_{\rm pair}({\bf r},{\bf r}; p)$ = $V_{o}$ + $(p-1) \Delta V$;
the pair--parameter $p$ takes account of the effect that
the interaction energy between a pair of particles in the same cell
depends on the number of different pairs present in the cell:
If there are 3 particles, and hence 3 pairs,
the first pair has an energy $V_{o}$, the second pair
an energy $V_{o} + \Delta V$, and the third pair an energy
$V_{o} + 2\Delta V$.
The total potential in cell {\bf r}
is then the sum of the pair potentials due to all particles in the
interaction neighbourhood
$N_{\rm int}({\bf r})$.

In order to minimize the number of free parameters of our model
we have set  $V_{\phi}$ $\equiv$ $V_{o}$ (equal to the interaction energy
of a single pair in a cell), and
$V_{e}$ $\equiv$ $V_{v}$ $\equiv$  $V_{1}$.
We are then left with 3 independent energy parameters
specifying the pair--interaction potential.
The orders of magnitude adopted for the latter are
$V_o$ = -3.0 MeV and $V_{1}$=-0.3 MeV;
  $\Delta V$ = +1.0 MeV.
The precise values are adjusted
to obtain optimal agreement with the observations, in particular to
secure the experimental mean binding energy per nucleon, of -8 MeV
in an intermediate--mass nucleus.
The value of $\Delta V$ has been estimated by requiring that 2 and 3
particles in a cell form a stable bound configuration; a larger
number of particles per cell leads to an unstable configuration.

Due to the discrete nature of the allowed CA states of motion,
a particle suffers a change of momentum which obeys Newton's
law of motion in a statistical sense only.
Two or more particles in a same cell undergo a scattering
which satisfies linear momentum conservation. The computational
details of the
treatment of the transitions among the particle states of motion
are given in \cite{AL99}.

\subsection{Fragmentation Clusters and Cluster Configurations}

A major aim of our simulation consists in constructing the
distribution law of the fragmentation clusters as actually registered by
real laboratory
detectors, in the form of the number of fragments,
$N$, against size, $a$ (and at fixed time $t$),
$N$ = $N(a,t)$.
In our experiments the `size' of a cluster is understood as
the number of particles in the cluster.
The  particles  are indiscernible,
so that any permutation among them which
does not alter the occupation of the individual cells, does
not produce a new cluster.

A fragmentation cluster of size $a$ of our CA context is
eventually identified
with a fragment of the laboratory nuclear fragmentation process.
It simulates a nucleus containing $a$ = $A$ nucleons.
The experimental counterpart
of our theoretical distribution, $N(a,t)$,
which ignores the charge of the fragments,
is then the distribution of the isobars as resulting
from the laboratory fragmentation process.

The precise specification of what we understand
by a `cluster' of size $a$ that
takes due account of the interactions included in our model
is given in the Appendix.
For the purposes of computing reaction temperatures,
we need to evaluate the number of distinct
configurations (`multiplicities') of the different
cluster--geometries compatible with a cluster of
specified size. This question is also dealt with in the Appendix.


\section{ The Set--up of the Numerical Experiments}

All of our experiments are carried out in the CM reference frame
whose origin coincides with cell $(0,0,0)$.

\subsection{ Initial Configuration}

The initial conditions for a dynamic run are as follows.
We simulate two nuclei, referred to as the `incident  nucleus'
$I$
and the `target'
$T$,
by two identical clusters  (in the sense of our definition) located in
the half--lattice $x < 0$ and $x > 0$ respectively.
The centres of mass of the clusters $I$ and $T$ are
required to lie on the $x$--axis. The mirror--symmetry
demands that we have
${\bf r}_{I}$ = $(-x_{M}, 0, 0)$ and
${\bf r}_{T}$ = $(+x_{M}, 0, 0)$ respectively.
The number of particles (and mass, in units of the
nucleon mass  $m$) of each cluster is
$A_T$ = $A_I$ = $A$ = 150.

Initially we assign each cluster cell a single particle,
so that the initial volumes of the clusters
are
$V$ = $A$ ${\lambda}^{3}$.
The shapes of these clusters are chosen to
approximate densely packed spheres.
This geometry leads to a radius--mass relation of the standard type
\begin{equation}
R \ \  = \ \ \left(\frac{3}{4 \pi}\right)^{1/3}  \ \ \lambda \ \   A^{1/3} .
\label{eq:2}
\end{equation}
Overall consistency with the empirical $R$--$A$ relation then
requires  $\lambda$ $\approx$ 1.9 fm.
The value we have  adopted for our model is $\lambda$ = 1.95 fm,
which produces the matter density  of a real nucleus of mass $A = 150$.
At time  $t = 0$ the two clusters
are just in contact (Fig. 1).

\subsection{ Initial Motion}

Dynamically, all particles
of each initial cluster, $I$ and $T$,  are in ordered  motion.
In the CM frame of $T+I$,
any particle of the target cluster has a velocity
${\bf v}_{T}$ = $-v {\bf e}_{x}$, while the velocity of
any particle in the incident cluster is
${\bf v}_{I}$ =  $-{\bf v}_{T}$ = $ +v {\bf e}_{x}$.
The ordered microscopic motion accounts for the initial macroscopic motion
of the
two clusters along the $x$--axis. The linear momentum of the global
system $T+I$  is zero in this frame.
The available impact energy is
\begin{equation}
E_{imp} \ \  = \ \ 2 \  \ A \ K\ ,
\ \ \  K \ = \ \frac{1}{2} m \ v^{2}. 
\label{eq:3}
\end{equation}
This energy is transformed into excitation energy
of the compound cluster; it leads eventually to the break--up of the latter.
(As in the traditional statistical models, we ignore here
particle--creation processes, in particular pion formation). 
The actual collision occurs at timestep $t=1$. 
Prior to the collision the model
describes two clusters $T$ and $I$ in uniform motion in the CM frame,
with opposite linear momenta ${\bf p}_T$ = $-A$ $m v $ ${\bf e}_{x}$
and ${\bf p}_I$ = $+ A$ $m v $ ${\bf e}_{x}$ = $-{\bf p}_T$.  The
velocities of the two clusters are opposite, of absolute value
\begin{equation}
|{\bf v}_T| \ \ = \ \ |{\bf v}_I|\ \ = \ \ v .
\label{eq:4}
\end{equation}

\subsection{ The Cluster Detectors}

In the  laboratory experiments
the detectors, which identify
and count the fragments, and measure their kinetic energies, are ideally
distributed isotropically around the collision centre.  We achieve an
acceptable approximation to isotropy respecting the lattice symmetry
of our CA environment  by placing our theoretical counting devices on
the 6 faces (CA lattice planes) of a cube of size $2 d_{D} + 1$
centred at the origin of the lattice (limiting planes defined by $x$ =
$\pm d_{D}$; $y$ = $\pm d_{D}$; $z$ = $\pm d_{D}$).  In conformity
with the laboratory experiments, the distance $d_{D}$ must be chosen
macroscopically large.  The minimum requirement is that $d_{D}$
exceeds the distance $d_{int}$ over which the fragments interact
(freeze--out radius).  For our typical cell size $\lambda$ and for
initial clusters of 150 nuclei each moving along the $x$--axis test
runs indicate that $d_{int,x}$ $\approx$ $d_{int,y}$ $\approx$
$d_{int,z}$ $\approx$ 20.  In our numerical experiments we have chosen
$d_{D}$ as large as computationally possible ($d_{D}$ $\approx$ 50 for
our CA universe of size $L$=127).

\section{ Fragment Identification and Counting Algorithm}

In models of statistical equilibrium, in which the fragments
are confined to a fixed finite volume $V_{B}$, the cluster
identification and count can be carried out with a standard
algorithm that consists in scanning the whole box available to the
fragments \cite{ST97}.
However, in the real laboratory experiment the detectors are not
uniformly distributed over a volume, so that the standard algorithm does not
duplicate the experimental procedure.
The fragment counting method we have set up  in our simulation
is devised to reproduce the principle of the laboratory counting procedure.

We isolate the `new' clusters, $\Delta N(a;t;x^{+})$, which
`pierce'  face $x = +d_{D}$ (our $x^{+}$ counter) of the
cube at timestep $t$.
The total number of clusters of size $a$,  $N(a;t;x^{+})$,
which have been traced by counter $x^{+}$ up to step $t$,
is then the sum of all `new'
clusters identified from timestep 1 up  to timestep $t$.
We have
\begin{equation}
\begin{split}
 N(a;t;x^{+}) \ \ 
 & = \  \ \Delta N(a;t;x^{+})\ \ + \ \  N(a;t-1;x^{+}) \\ 
 & = \ \Delta N(a;t;x^{+}) \  +  \ \Delta N(a;t-1;x^{+})\ +\ldots
+ \Delta N(a;1;x^{+}) \ . 
\end{split}
\label{eq:5}
\end{equation}

The identification at the other counters
$x^{-}$, (face $x = -d_{D}$);
$y^{+}$,
$y^{-}$, (faces $y = \pm d_{D}$);
$z^{+}$,
$z^{-}$, (faces $z = \pm d_{D}$),
follows the same scheme.
The total number of all distinct clusters $N(a;t)$ traced up to time $t$
is the sum over the measurements of all 6 detectors.

The relative distribution of the fragments, $Pr(a)$, probability of a
fragment of size $a$, is given by
\begin{equation}
Pr(a) \ = \ \lim_{t \to \infty}\ \frac{N(a;t)}{N(t)},\ \text{with} \
N(t) \ \equiv \ {\sum_{a=1}^{2 A}  N(a;t)}  . 
\label{eq:6}
\end{equation}

Finally, if we repeat the same
simulation
a large number $r$ of times, we have
$$
\langle Pr(a)\rangle \ = \
\frac{\sum_{c=1}^{r}  {N}^{(c)}(a;t) }{\sum_{c=1}^{r}  {N}^{(c)}(t) }
\ = \
\frac{\sum_{c=1}^{r}  {N}^{(c)}(t) {Pr}^{(c)}(a) }
{\sum_{c=1}^{r}  {N}^{(c)}(t) } .
\eqno (6.a)
$$
where ${N}^{(c)}(t)$ is the total number of fragments of all  sizes,
${N}^{(c)}(a;t) $ is the fragment distribution,
and ${Pr}^{(c)}(a)$ is the relative fragment distribution in
the $c^{th}$ experiment.
If the number $r$ of experiments is chosen large enough, a
stable and well defined distribution is expected to emerge.

To find $\langle Pr(a)\rangle$ we have developed a straighforward
algorithm for the identification
of fragmentation clusters of size $a$ which enter the plane
$x = +d_{D}$ at step $t$. The algorithm exploits the
property that since a free fragment propagates with
speed $v$, $\Delta N$ represents the number of fragmentation
clusters which intersect the plane  $x = +d_{D}$,
but which do not intersect the plane $x = +d_{D}+1$.

To the extent that our CA model incorporates an acceptable
approximation to the physics of the
fragmentation process, the
relative fragment distribution as derived from our simulations
should duplicate the laboratory distribution
${Pr}^{(lab)}(a)$ as measured by real detectors.
In fact, consider a laboratory run of total duration $t_{exp}$,
operating under stationary conditions.
We then have a constant flux of infalling particles
colliding with a flux of  target particles. Under
typical operating conditions these collision
processes are binary collisions (each individual collision process
occurs independently of the other collision processes).
Collision   $c$ therefore produces in the array of detectors
 a distribution given by Eq. (6).
 If  $r(t_{exp})$ is the total number of collisions which
occur over the whole run of the experiment,
then the detectors register
the following relative  fragment distribution
$${Pr}^{(lab)}(a) \ = \
\frac{\sum_{c=1}^{r(t_{exp})}  {N}^{(c)}(a;t) }
{\sum_{c=1}^{r(t_{exp})}  {N}^{(c)}(t) }
\ = \
\frac{\sum_{c=1}^{r(t_{exp})}  {N}^{(c)}(t) {Pr}^{(lab,c)}(a) }
{\sum_{c=1}^{r(t_{exp})}  {N}^{(c)}(t) }.
\eqno (6.b) $$
The notations adopted in these expressions are essentially
the same as in the theoretical case
(${N}^{(c)}(a;t)$, ${N}^{(c)}(t)$: fragment numbers
in collision $c$; ${Pr}^{(lab,c)}(a)$:
 relative fragment distribution of the
$c^{th}$  collision; ${Pr}^{(lab)}(a)$: final
average relative experimental distribution of the fragments).


\section{ Dynamic Results}

The evolution of the particles of the two colliding clusters
is followed with our CA program over a total time--interval
$t_{max}$ not exceeding $(L-1) - (d_{int}+d_{D})$
(about 70 in our experiments).
The order of this time interval is fixed by the observation
that it is the shortest time it takes an individual particle, or a cluster,
ejected in the collision, to migrate
through the available CA lattice,
to be  reflected on the boundary of the lattice universe,
and to be sent back to a detector plane.
For times $t>t_{max}$ a reflected fragment
could collide with outflowing fragments;
this would violate our assumption of
non--interaction of the fragments at distances exceeding $d_{int}$.

We begin with a discussion of a first physically realistic effect.

(i) At the collision site we observe a central concentration of
nucleons (a compound nucleus) of size  $A_C(t)$.
This concentration progressively loses
individual particles  as well as fragments.
The sequence of frames (a), (b), (c), (d) exhibited in Fig. 2
and corresponding to an impact energy per nucleon
$E_{imp}/A$ = 3.75 MeV, illustrates this situation
in greater detail.
Frame (a) ($t$ = 20) indicates that besides isolated particles
leaving the centre
two large fragments were symmetrically ejected along the
collision line; smaller fragments were blown along the $y$-- and
$z$--axes normal to the collision line ($x$--axis).  The collision
site remains a high density zone, which is clearly visible on the
later frames (b), (c), (d) ($t$ = 40, 50, and 60).  An evaporation of
small fragments and individual particles continues from this central
condensation zone, at a rate which is expected to depend on the impact
energy, $E_{imp}$ (the only free parameter of our series of
experiments).

The experiment seems to suggest that the collision phenomenon is actually a
two--stage process. In the first phase of the collision,
the nucleus is broken up into several large fragments; the latter are
the larger the higher the impact energy is; for a high enough impact
energy, only 2 fragments emerge from the collision (negligible
interaction between the two nuclei).  The second phase, starting at some
timestep $t_{o}$, is a more gentle
evaporation from a central fragment.

To quantify the  phenomenon of emission of clusters
we fix a small reference volume
(a cube of 21$^{3}$ cells) which encloses the compound nucleus,
and which we regard as a rough approximation to the space occupied by the
latter.  It is then reasonable to assume that the rate of particle loss
from this reference volume is proportional to some a priori unknown power
$\delta$ of an excess number of particles over an equilibrium number,
$A_{o}$, in the reference volume, or
\begin{equation}
d/dt \ A_C \ = \ -\  \beta (E_{imp}) \ (A_C -A_{o})^{\delta}\ ,
\ \ \beta (E_{imp}) \  =    \  \gamma \ {E_{imp}}^{1/2} .
\label{eq:7}
\end{equation}
The functional form of   $\beta (E_{imp})$ follows
from a dimensional scaling argument.  In our
series of experiments in which the impact energy alone is regarded as
a parameter, all other free coefficients being held fixed, we can
express mass in terms of the nucleon mass, and volume in the reference
volume, so that mass and volume (and hence length) become
dimensionless; in this system the impact energy has the dimension of
an inverse time squared.  Hence the above relationship.
The remaining
factor $\gamma$ depends on the model constants (i.e., reference volume,
mass of nucleons, nuclear interaction constants, etc).

Integration of Eq.  (7), from the moment $t_{o}$, at which evaporation
starts, to the current time $t$, yields
$$ (A_C(t) - A_{o})^{1-\delta} \ =
\  (A_{C}(t_{o}) - A_{o})^{{1-\delta}} \ -
\ (1-\delta) \ (t-t_{o}) \ {E_{imp}}^{1/2}\ \gamma,
 \ \ {\rm if}\ \  \delta \ne 1,\eqno (7.a)$$

$${A_C(t)\ -\  A_{o}}\ = \ (A_C(t_{o})-A_{o})\ \exp[-{E_{imp}}^{1/2}\
\gamma\ (t-t_{o})],
 \ \  {\rm if}\ \  \delta = 1. \eqno (7.b)$$
Fig.  3 shows the curves $(A_C(t) - A_{o})^k/(A_C(t_{o}) - A_{o})^k$
for values of $k$ (=1-
$\delta$) ranging from 0.1 to 0.5 (fine lines), as well as $\ln [A_C(t) -
A_{o}]/ \ln [A_C(t) - A_{o}]$ (heavy line); time is measured with
$t_{o}$ taken as origin.  The parameters $t_{o}$ and $A_{o}$ are read
off from the numerical run ($t_{o}$ = 11; $A_{o}$ = 36 in the case of
the experiment exhibited in Fig.  3).  The curves in Fig.  3, which
are strongly nonlinear for larger values of $k$, tend to become linear
in the limit $k \to$ 0, as is found from a linear regression analysis
(over the 25 timesteps shown). Hence Eq.  (7.b) ($\delta$ = 1) gives the
best fit.  In other words, the excited compound cluster essentially
suffers a standard disintegration.  The decay time for the specific
experiment we have displayed is $\tau_{dis}$ = 6.58.

(ii) A second conspicuous feature of the plot of
our numerical results
is an artefact of the CA lattice symmetry and
computational procedure.
In each lattice direction, $\pm {\bf e}_{i}$, $i=x,y,z,$
we observe a column of particles or clusters that propagate all
with maximum speed $v$ (Eq. 1) away from the collision
site.

Fig. 2 discloses that the density of particles in each column
increases with distance, with a peak density at
the two ends of each column.
The peak density corresponds to the effect of the violent break--up
occurring in the first phase of the collision.  The subsequent gentle
decrease in time of the content of the residual compound cluster, $A_C(t)$,
implies that the rate of evaporation (Eq.  7) decreases with time as
well.  Therefore the density in the columns decreases as we approach
the central compound cluster; at the latter central cluster, the
common centre of the columns, a density maximum survives.

Statistically the particle distribution at any time
$t$ preserves the symmetry of the initial configuration.
Fig. 2 demonstrates indeed that the pattern
possesses the following  symmetry elements:
The $x$-- axis is a fourfold  axis; the planes
$(x,y)$ and $(x,z)$, as well as their bisectors are
reflection planes;
the $y$ and $z$--axes are binary
axes, and so are the two diagonal axes;
$(y,z)$ is a reflection plane.
The invariance group of this statistical configuration
is $D_{4h}$ (in standard notations for the point--groups;
cf Landau--Lifshits, Quantum Mechanics).
The distribution of the particles along the collision axis $x$
is clearly seen to be different from
the distributions along the $y$ or $z$--axes,
(cf in particular the large fragments on the $x$ axis).
The collision axis remains
a privileged direction all over the experiment.

At timestep $t$ = 0, the initial conditions of the collision set--up
in our CA lattice environment, Fig. 1,
create a  symmetry breaking from the original octahedral (plus translational)
symmetry of the empty (infinite) lattice (the  `vacuum state'), $O_{h}$,
to the symmetry  $D_{4h}$ of our initial configuration.
At all later times, $t > 0$, our CA results
demonstrate that the latter symmetry
is statistically preserved (cf Fig.  2). The collision process itself,
starting at $t$ = 1, induces no overall geometric symmetry transition.
The occurrence of such a transition would be direct evidence for a
second order phase transition.  If a second order phase transition
does occur in the fragmentation process, then it must be related to a
finer symmetry breaking not immediately manifest in the spatial
distribution of the fragments.

The initial conditions for a head--on collision
of two identical nuclei occurring in the continuous physical configuration
space
break the symmetry of the original `vacuum' (spherical plus
translational symmetry) transforming it into a cylindrical symmetry, of
axis coinciding with the collision axis, $K_{h} \to D_{\infty h}$.
Provided that our discrete CA model can capture the essential physics
of the real fragmentation mechanism, the results of the above
numerical experiments are indicative that the full cylindrical
symmetry should be preserved in the real laboratory experiment.  This
symmetry is indeed consistent with the $4 \pi$ detector measurements.

Our next task is an attempt at wiping out the CA artefact (ii)
in the computed spatial distribution and to convert the latter
into a form more directly comparable with the laboratory experiments.

We reconstruct a
$D_{\infty h}$--symmetric
number distribution, $\rho^{rec}({\bf r};t)$,  from the CA
$D_{4h}$--symmetric
number distribution, $\rho^{CA}({\bf r},t)$,
by the following procedure. We expand the new distribution in
Legendre polynomials
\begin{equation}
\rho^{rec}({\bf r};t) \ =\ \sum_{n=0}^{\infty}\ \rho_{n}(r;t) \
P_{n}(\cos \theta),
\end{equation}
where the angle $\theta$ is the angle between the position vector
${\bf r}$ and the collision axis (oriented from the $I$ to the $T$
nucleus).
Since the CA experiments indicate that the difference in the distribution
along the collision axis and the axes normal to the collision axis
is relatively small, we truncate the expansion after the dipole terms.
In this lowest order approximation we then write
$$\rho^{rec}({\bf r};t) \ = \ \rho_{c}(r;t)\   +\  \eta (r;t)\ \cos \theta
. \eqno (8.a)$$
The functions $\rho_{c}(r;t)$ and $ \eta (r;t)$ are then
determined from the CA distribution by
$$ \rho_{c} (r;t) \ = \ B\  \langle \rho^{CA}({\bf r};t) \rangle_{y,z}, $$
$$ \rho_{c} (r;t)\ + \ \eta (r;t)\ =
\  B\ \langle \rho^{CA}(0,r,0;t)  \rangle_{x}, \eqno (8.b)$$
where the righthand sides are averages over the positions
$r$ = $|y|$ and $|z|$ on the $y$  and $z$ axes,
and $r$ = $|x|$  on the $x$ axis respectively;
$B$ is a normalisation constant which is
fixed by requiring that the integration of the
distribution over the available space
(the sum of $\rho^{rec}({\bf r};t)$ over all CA cells ${\bf r}$)
is equal to the total number of nucleons (=300 in our experiments).

Panel  (b$'$)  of Fig. 2 shows a stereoscopic plot of
the symmetrized space distribution (8) reconstructed from
the CA distribution displayed in panel (b).  The plot is generated by a
straightforward Monte--Carlo procedure distributing 300 particles in
conformity with the statistical law (8.a).  We observe that this
distribution falls off with distance $r$ from the centre (with an
exponent $<2$), except that the higher density of the distribution
front survives.  We should point out that the method just redistributes the
individual particles without conserving  the clusters.

(iii) The cluster--size distribution as generated by our CA dynamics
and registered by our counters $x^{+}$,etc, obeys a power law.  This result is
illustrated in greater detail in Fig.  4 , which exhibits two instances,
one at very low impact energy, $E_{imp}/A$ = 0.307 MeV (panel a), and
the other at a 13 times higher energy $E_{imp}/A$ =3.973 MeV (panel
b).  The plot shows (a non--normalised form of) relation (6) in log--log
format, obtained from a total number of runs $r$ = 16.  Over
the range $a$ $<$ 10, both curves appear as straight lines, with
negative slopes $\tau$ = 2.56 (low energy) and 2.64 (high energy).
Our experiments indicate that this property holds at least up to
$E_{imp}/A$ $\approx$ 10 MeV, where $\tau$ takes the value 2.65.
We are entitled to conclude that in the range of impact energies
$E_{imp}/A$ $<$ 10 MeV, the (smaller) fragments obey a power law with
negative slope $\tau$ $\approx$ 2.6, which is independent of the
energy.  As the impact energy increases larger clusters are being formed; the
tail of the distribution then tends to become longer.

Critical exponents $\tau$ of
2.64 and 2.65 have been measured in the laboratory, for the fragment
distribution of proton--Kr and proton--Xe fragmentation.
We should mention also that in the field of ion cluster fragmentation,
fragment distributions obeying power laws of exponents of 2.56 and
2.63 \cite{FA97} have been isolated.

It appears that the slope values  of the fragment distribution
we obtain from our dynamic results are in better agreement with the
laboratory data than the values derived from standard statistical
theories, which are close to the value given by mean field theory
(\cite{JA83}: $\tau$ = 2.33).  This remains also true for ion cluster
fragmentation problems, even though our model parameters were not
adjusted to that particular situation.

\section{ Formal Temperatures. Comparison with Laboratory Experiments }

With the exception of the symmetrised distribution (8),
the results described in the previous Section
are direct results of our $N$--body simulations.
They involve no extra approximations
besides the assumptions inherent in any CA modellisation
(space and time discretisation) and the schematisation of the
interaction potential among particles.  In particular, the dynamic
calculations dispense with the hypothesis of formation of a compound
cluster in the collision process.  Our numerical experiments indicate that
a configuration showing a higher central concentration, which we refer
to as the compound cluster, always emerges in the simulation (Fig.  2),
provided only that the collision energy is low enough as compared to
the binding energy of the nucleons in the nuclei.  The compound
cluster (i.e. high central concentration) survives over a period of time
which can be estimated from Eqs.  (7).

Within this compound cluster, and, in principle, also within our CA universe
$U$ (127$^{3}$ cells; toroidal topology), a variety of statistical
equilibria $e$, $e'$, $e''$, $\ldots$ are conceivable among the
different components.

(a) If we identify the components as the individual particles
(nucleons in our specific nuclear fragmentation simulation; etc) of a
configuration $Z$, then we may have a statistical equilibrium in $Z$,
${e_{E}}^{n(Z)}$, due to energy exchange ($E$) among the particles
($n(Z)$) of configuration $Z$.  The configuration $Z$ may be the
compound cluster $C$, or our entire CA universe $U$.

(b) If we identify the components as the clusters of particles $c$ (which in
turn, when evaporating from the compound cluster become the observable
fragments), then we may have (i) a statistical equilibrium due to
energy exchange among the clusters of a configuration $Z$, $c(Z)$,
${e_{E}}^{c(Z)}$.  But we may also have (ii) an equilibrium due to
exchange of energy as well as particles ($P$) among
these clusters, ${e_{E,P}}^{c(Z)}$.  Under the latter alternative, the
observed distribution of fragments against size will be the thermal
reaction--equilibrium distribution of the clusters.  For the sake of
completeness we mention further (iii) that within each individual
cluster $c_{f}$, the component particles may exchange energy among each other;
this process may then lead to a special type of equilibrium,
${e_{E}}^{n(c_{f})}$, inside each cluster $c_{f}$.

We define here a temperature parameter $T(e)$, which is associated with a
specific thermodynamic equilibrium process $e$, as follows : The
parameter $T(e)$ is the thermodynamic temperature that reproduces the
macroscopic properties of the specific equilibrium $e$ assuming it is
realized.  If $\omega$ = $F(e;T)$ is the thermodynamic relation that
expresses the observable $\omega$ as a function of the thermodynamic
temperature $T$ under equilibrium conditions $e$ (other thermodynamic
variables being held fixed), then from the measurement of $\omega $
and the thermodynamic relation we set $T(e)$ $\equiv$ $T$.  (Note that
our way of introducing temperatures differs from the more formal procedure
adopted in Chernomeretz et al.  \cite{CH01}).  We are free to use this
thermodynamic relation in a formal way to determine a parameter $T(e)$
characterising a given macroscopic configuration, whether or not the
specific equilibrium $e$ holds.  We begin with listing the
thermodynamic relations of relevance for our purposes.

In principle, all of the above--listed equilibria, $e$, $e'$, $e''$,
$\ldots$ may arise in our numerical experiments,
or in the laboratory experiments
(in the sense that there is for instance no membrane surrounding
a fragment that would prevent the exchange of particles, etc).  The question
is rather: Given an equilibrium process $e$, does the collision system
we investigate survive over a time--span that is long enough for the
equilibrium $e$ to establish itself?

If the collision system reaches a full thermodynamic equilibrium,
then we must have
\begin{equation}
T(e) \ = \ T(e') \ = \  \ldots \ \equiv  \ T \ ,
\label{eq:9}
\end{equation}
where the formal temperatures $T(e)$, $\ldots$ obtained as
indicated;  all of these formal parameters are then equal;  they define
the single thermodynamic temperature $T$.  Conversely, if relation (9)
is violated, then the relevant statistical equilibria are not
realized.  We have indeed good reasons to believe that some of the
possible equilibria listed above will never materialize in our system
(cf below); or, alternatively, some equilibria cannot establish
themselves over certain ranges of the impact energy.  Our numerical
results confirm this point.

Even if a state of full thermal equilibrium is not attained for
the system we are investigating,
the formal
temperatures, $T(e)$, $T(e')$, $\ldots$,
may remain perfectly useful parameters
for the purposes of comparison of the numerical results
with actual laboratory experiments in the following sense.
Suppose a given observable $\Omega$, measured in the laboratory,
is plotted
against the temperature  parameter, $T(e)$,
$\Omega$ = $Q(T(e))$,
this temperature  being measured
according to a well--defined protocol (cf above).  A typical instance is
provided by the (formal) caloric curve, $T({e_{E,P}}^{c(C)})$ {\it vs}
$E_{imp}$, (equilibrium process: energy and particle exchange among
clusters in the compound cluster).  The necessary condition for a
given theoretical model, such as our present CA model, to be an
adequate model for the fragmentation process, is then that this model
can duplicate the experimental $\Omega$ = $Q(T(e))$ relation.  The
agreement must hold provided that the formal parameter $T(e)$ be obtained
in conformity with the experimental protocol.  It should be kept in
mind that there is no guarantee that an easily measurable temperature
parameter of our CA gas (such as $T({e_{E}}^{n(U)})$ can be
substituted to the actual experimental parameter (temperature
associated with ratio of yields $T({e_{E,P}}^{c(C)})$ ).

We have measured the following collection of temperature parameters for our
CA system (adapted to nuclear fragmentation).

\subsection{ Nucleon-gas temperatures:
Global nucleon--gas temperature, $T_{n(U)}(t)$, and
nucleon--gas temperature in compound nucleus, $T_{n(C)}(t)$}

The temperature parameters $T_{n(U)}(t)$ and $T_{n(C)}(t)$
are the formal temperatures of a gas of nucleons in
thermal equilibrium under exchange of energy in the
CA universe $U$ and in the compound nucleus $C$
respectively. With the above notations
$T_{n(U)}(t)$ $\equiv$ $T({e_{E}}^{n(U)};t)$,
$T_{n(C)}(t)$ $\equiv$ $T({e_{E}}^{n(C)};t)$;
the extra argument $t$ indicates that these
parameters depend on time.

The total number of nucleons of our system,
the CA universe $U$, is conserved, $A_{U}(t)$ $\equiv$ $A$ (=300).
The number of nucleons in the
compound nucleus $C$ at timestep $t$, $A_C(t)$, is variable,
and so are
the number of nucleons in motion in the universe,
$A_{m(U)}(t)$,
and the number of nucleons in motion
in the compound nucleus,
$A_{m(C)}(t)$.
The temperature parameters, which are
measures of the average kinetic energy per nucleon
of the gas of nucleons in $U$ and $C$ respectively,
are then given in terms of these numbers of nucleons
by the following expressions
\begin{equation}
\frac{3}{2}\ \ k \ T_{n(U)}(t) \ =
\ \frac{1}{2 \ A} \ m \ \sum_{\alpha=1}^{A} ({\bf v}_{\alpha}(t))^{2} \ =
\ \frac{A_{m(U)}(t)}{A} \ K\ \  < \ \  K, 
\label{eq:10}
\end{equation}
\begin{equation}
\frac{3}{2}\ \ k \ T_{n(C)}(t) \ =
\ \frac{1}{2 \ A_C(t)} \ m \ \sum_{\alpha=1}^{A_C(t)} ({\bf
v}_{\alpha}(t))^{2} \ =
\  \frac{A_{m(C)}(t)}{A_C(t)} \ K\ \  < \ \ K.
\label{eq:11}
\end{equation}

In these expressions $k$ is the Boltzmann constant,
${\bf v}_{\alpha}(t)$ is the velocity
of the nucleon labelled ${\alpha}$ and taken at timestep $t$, and $K$ is the
kinetic energy of a moving CA particle (Eq.  3).  Note that all formal
temperature parameters we shall introduce are time--dependent.

The second temperature parameter, $T_{n(C)}$ (Eq. 11),
 measures the physically meaningful average kinetic energy of a gas of
particles in the compound cluster,  which can in
principle reach a thermal equilibrium in our CA model, or in the
laboratory; hence it can represent a genuine temperature of a gas of
nucleons.  This is the case if the energy exchanges among the nucleons
have time to establish themselves, i.e., if the lifetime of the
central condensation (cf Eqs.  7) exceeds the average collision time
among nucleons in $C$ (a few timesteps).

Figure 5 shows a caloric curve
$T_{n(C)}$  against the impact energy per nucleon $E_{imp}/A$,
with  $T_{n(C)}$ $\equiv$ $T_{n(C)}(t_{ref})$ computed from our CA runs
at a reference timestep $t_{ref}$ (= 17 for reasons to be discussed
below).
In the low energy range,
up to $E_{c}/A$ $\approx$ 7--8 MeV/nucleon, the temperature parameter
rises slowly with impact energy, approximately as
$$T_{n(C)}\ = \ 0.15 \ E_{imp}/A,\ \    E_{imp}\ <\ E_{c}  .\eqno (11.a)$$
In the high energy range the rise is steeper, and we approach
$$T_{n(C)}\ = \ \frac{2}{3} \ (E_{imp}-E_{c})/A ,
\ \    E_{imp}\
\gg \ E_{c}   .\eqno (11.b)$$
The slope 2/3 in the latter relation is indicative
that the random kinetic energy of
the nucleons is asymptotically equal to the impact energy minus
$E_{c}$. Accordingly this critical energy appears as a binding
energy of the nucleons in the compound nucleus. The numerical
value of $E_{c}/A$, which is of the order of the average binding
energy per nucleon in a stable nucleus,
is indeed consistent with this interpretation.

As can be seen in Fig. 5, the slope of the asymptotic expression
(11.b) is in line with the high energy branch of the
experimental caloric curve of the  $^{197}$Au--$^{197}$Au
fragmentation  process \cite{NU97}.
The NuPECC  caloric curve exhibits a
plateau, at a temperature level of 4.5--5.0 MeV,
which is not  reproduced by the CA nucleon--gas temperature;
however, the change of slope in the theoretical curve
coincides with the transition from a plateau to
a rising behaviour. We should stress that the experimental points of
Trautmann's analysis, also plotted in Fig. 5, indicate
no proper plateau; they rather follow a rising curve
similar to the CA $T_{n(C)}$ curve.
Quantitatively the experimental data (NuPECC or Trautmann) are shifted by
$\Delta T$ $\approx$  3.5 Mev above the CA curve.

The first parameter, $T_{n(U)}(t)$, must be regarded
as an artificial temperature. It measures
an average kinetic energy of nucleons composing
a mixture of two gases, which in the context of
our CA model, or the real laboratory experiments, never interact.
Namely we have on the one hand
a gas of nucleons essentially trapped inside the compound nucleus $C$, and
on the other hand
an expanding gas outside the compound nucleus, $U-C$. In the latter gas
the collisions are negligible; no energy
exchange can take place among the outer nucleons. All of these latter nucleons
conserve the momentum and kinetic energy (maximum energy $K$)
they have acquired at the
moment they evaporate from $C$.
To simulate realistic laboratory
nuclear fragmentation, our CA experiments must consist in
relatively short runs, of a total number of timesteps
$t_{max}$ essentially chosen as follows (cf Section 4).
Once a fragment has left the collision site it suffers no
further interactions.
In the finite CA universe of our model this requires that an ejected
fragment be not allowed to be reflected on the boundaries of the cubic
CA universe, and sent back to the reaction site.  Very roughly we then
choose $t_{max}$ $<$ $L/2$.  (Interactions in $U-C$ would require that
the number of timesteps obey $t_{max} \gg L$).

Even though $T_{n(U)}(t)$ has a formal meaning only
in the nuclear fragmentation problem,
it can be used to find the physically significant
temperature parameter of the compound
nucleus, $T_{n(C)}(t)$.
In fact, in the numerical experiments we can easily
compute $T_{n(U)}(t)$; we  can also easily count the outer nucleons,
$A_{U-C}(t)$. Taking then account of the relations
$$A\ = \ A_{C}(t)\ + \ A_{U-C}(t),\ \ {\rm and} \ \
A_{m(U)}(t)\ = \ A_{m(C)}(t)\ + \ A_{U-C}(t), $$
we obtain from Eqs. (10,11)

\begin{equation}
T_{n(C)}(t) \ = \  T_{n(U)}(t) \ -  \left(  \frac{2}{3} \ \frac{K}{k}
\ -\ T_{n(U)}(t) \right) \
\frac{A_{U-C}(t)}{A - A_{U-C}(t)}.
\label{eq:12}
\end{equation}
This relation demonstrates
that the temperature  parameter in the compound nucleus at timestep $t$,
$T_{n(C)}(t)$, is always less than the
formal temperature parameter of the CA universe,  $T_{n(U)}(t)$,
taken at the same timestep $t$.
Fig. 5 exhibits also the formal caloric curve
$T_{n(U)}(t)$  {\it vs}  $E_{imp}/A$,
at timestep $t$ = $t_{ref}$ (= 17),
which is seen to obey the inequality
$T_{n(U)}(t)$ $>$ $T_{n(C)}(t)$.
In the lower energy range the two curves
$T_{n(U)}(t)$ and $T_{n(C)}(t)$ are nearly superposed,
indicating that at the reference time few nucleons have escaped the compound
nucleus. Even at the highest impact energies
we have investigated,
the difference in the two formal temperatures
does not exceed 1.5 MeV.

\subsection{ Cluster-gas temperatures $T_{c(U)}(t)$
and  $T_{c(C)}(t)$}

At timestep $t$ consider the specific $f^{\rm th}$
fragmentation cluster, $c_{f}$, $f$ = 1,2,$\ldots$,$F_{Z}(t)$,
a representative of the cluster equivalence class  $C(a_{f})$.
 $F_{Z}$ denotes the total number
of clusters in the configuration  ($U$, or $C$).
Denote by $\langle {\bf v}_{f}(t)  \rangle$  the velocity of this
cluster (average of the velocities of the component nucleons)
at step $t$.
Then the cluster--gas temperature is the temperature parameter
associated with the equilibrium brought about by the exchange of
energy among the clusters in a given configuration $Z$.
We have
$T_{c(Z)}(t)$ $\equiv$  $T({e_{E}}^{c(Z)};t)$, $Z$=$(U,C)$.
Hence

\begin{equation}
\frac{3}{2}\ \ k \ T_{c(Z)}(t) \ = \
\frac{1}{2 \ F_{Z}(t)} \ m \  \sum_{f =1}^{F_{Z}(t)} \ a_{f}
(  \langle {\bf v}_{f}(t) \rangle)^{2}  .
\label{eq:13}
\end{equation}

As in the case of the nucleon--gas, the
cluster--gas temperature parameter $T_{c(U)}(t)$ is again
a formal magnitude, since the cluster--gas outside the
compound nucleus does not interact within
our runs of $t_{max}$ timesteps.
On the other hand the clusters in the compound nucleus $C$
may have time to thermalise, so that the temperature
parameter $T_{c(C)}(t)$ does represent a physically meaningful
temperature.

On Fig. 5 we have superposed  the formal theoretical caloric curves
$T_{c(U)}(t)$ and $T_{c(C)}(t)$ {\it vs} impact energy $E_{imp}/A$
(at timestep  $t_{ref}$).
We note that the inequality between the two temperature parameters
of the nucleon gas in $U$ and $C$, read off from Eq. (12),
is preserved in the case of the
cluster gas in $U$ and $C$ :
$T_{c(U)}(t)$ $>$ $T_{c(C)}(t)$
(outer fragments move with maximum speed).
The cluster--gas curves have
much steeper slopes than the nucleon--gas curves,
implying the further inequality
$T_{c(C)}(t)$ $>$ $T_{n(C)}(t)$.

The higher cluster--gas temperature reflects the following property.
In a given configuration $Z$ the
number of clusters is smaller than the number of nucleons
while the total energy to be shared among clusters, or among
nucleons
is the same. Therefore the average energy per cluster is
larger than the average energy per nucleon.
At high impact energies,
$E_{imp}/A$ $>$ average binding energy
per nucleon in a stable nucleus,
one might expect intuitively
that the physically meaningful caloric curve for a fragmentation
process should be the nucleon--gas curve $T_{n(C)}$.
The  high energy available, when shared among the nucleons,
would indeed produce average energies per nucleon which exceed the binding
energies of any fragment.
However, our CA experiments demonstrate that
in the case of high impact energy
the merged $T$ and $I$ nuclei break
immediately up into essentially 7 large fragments. Two
compact fragments carrying a sizeable fraction of the mass
of the $I$ and $T$ nuclei, continue to propagate along the collision
 axis.  Four smaller fragments are ejected along the
$y$-- and $z$--axes respectively. The residual nucleons form a
concentration of matter at the  centre. Only the latter can
play the part of a compound nucleus
in which actual energy sharing  may occur.
Qualitatively this initial break--up  survives
at lower energies, except that the 6 ejected fragments become
progressively smaller while the central residual becomes larger
as the impact energy is decreased (cf the sequence shown in Fig. 2
for an impact energy 3.75 MeV per nucleon, where the 7 fragments
are clearly distinguishable on all panels).

The observed  scenario is indicative that
at high impact energies it is the
cluster--gas temperature that supplies the physically
meaningful characterization of the laboratory
caloric curve. This temperature takes properly care of the
contribution of the large fragments in the energy balance.
Any laboratory measurement technique,
(and any theoretical procedure) of
a temperature assignment
ignoring the largest clusters must fail to provide a
physically relevant temperature estimate.

This point is of importance in connection with the
reaction temperatures. The latter, to the extent that
they typically refer to fragments of low size (cf below),
are not physically representative in the high energy range.

\subsection{ Reaction temperatures, $T_{cc'\ldots/ c''\ \ldots}(t)$}

The favoured laboratory method for assigning an experimental temperature
to a fragmenting nuclear system consists in measuring ratios of yields
of different fragments, assuming a statistical equilibrium of
general type
 ${e_{E,P}}^{c(C)}$
among the fragments \cite{CA94}.
Examples are the ratios (${^3}$He/${^4}$He) and (${^6}$Li/${^7}$Li)
for the earlier analysis of the  ALADIN experiments \cite{PO95},
 \cite{AL85} and \cite{CA94}, and other ratios of populations of
isotopes in the more recent analysis
of the same experiment \cite{NU97}; (cf also \cite{PO87}, \cite{KU91}).
In \cite{HA96} the ratios
(${^3}$He/${^4}$He)/(d/t) and (${^3}$He/${^4}$He)/(${^6}$Li/${^7}$Li)
are considered for the multifragmentation resulting from
an Au target bombarded by C ions. Several other instances of measured ratios
are listed in \cite{RI01}.
In all cases
populations of light nuclei alone have been investigated, from which a
specific temperature parameter is derived via the standard relation for
chemical equilibria (cf \cite{LA55};  the specific
nuclear context is dealt with in \cite{BO00}).

In the framework of our CA formulation which ignores electric charge, the
chemical equilibria among different fragments are not directly comparable
with the real nuclear equilibria investigated, which involve isotopes.
Under conditions of true thermal equilibria, the ratio
among any group of isotopes, or of isobars, is governed by the same
thermodynamic temperature.  But if the system is not in a genuine
state of equilibrium under nucleon exchange, then the temperature
parameter is specific for the precise reaction process.  Different
reactions, and hence different measured ratios of fragments in the CA
experiments, lead to different temperature parameters.  The same
conclusion holds for the laboratory ratio measurements.

In fact, as transpires from \cite{NU97},
different experimental ratios, and different analyses of these ratios,
have led to different shapes of the caloric curve in the case of the
$^{197}$Au--$^{197}$Au fragmentation process. In Trautmann \cite{NU97}
a rising pattern for the He--Li ratio is identified, while a nearly
constant temperature is found for other ratios; moreover, the earlier
NuPECC caloric curve \cite{NU97}, based on other ratios, exhibits a
temperature--plateau (not present in Trautmann), approximately over
the range 3 to 10 MeV per nucleon.
In Fig. 5 we have superposed the various available experimental points
defining caloric curves for the symmetric $^{197}$Au--$^{197}$Au fragmentation.
This laboratory process comes close to our CA simulation,
even though the total number of nucleons involved in the laboratory
is  4/3 times the  number of nucleons of our  simulation.

To formulate the relevant statistical expressions for reaction equilibria
in the CA context, consider the equilibrium among
the cluster classes
$C(a_{f})$, $C(a_{f'})$, $\ldots$,
described by the stoichiometric scheme
\begin{equation}
\nu_{f}\ C(a_{f})  \ +\ \nu_{f'}\ C(a_{f'})\ + \  \ldots  \
\rightleftharpoons\
\nu_{f''}\ C(a_{f''}) \ + \ \ldots, 
\label{eq:14}
\end{equation}
($\nu_{f}$, $\nu_{f'}$,$\ldots$ integers consistent with conservation of
the number of nucleons in the reaction process : $\nu_{f} a_{f}$ + $\nu_{f'}
a_{f'}$ + $ \ldots$ = $\nu_{f''} a_{f''}$ + $\ldots$)

Denote by $N(a_{f},t)$ the total number of clusters
of class $C(a_{f})$  present in our system at  timestep $t$.
If a statistical equilibrium holds, then
the standard statistical procedure allows us to write
\begin{equation}
N(a_{f},t) \ = \ \frac{ V (2 \pi m  k T(t) )^{3/2} }{h^3}
\times \frac{ {a_{f}}^{3/2} }{ a_{f}! } \times
 \ \left( \  \sum_{j=1}^{J}  \ g_{j} \ {\exp} (-E_{int,j}/kT(t) ) \  \right)
\times   \exp (\lambda \nu_{f}\ a_{f}).
\label{eq:15}
\end{equation}
In this expression
$V$  denotes the reaction volume; $T(t)$ is the equilibrium
temperature at timestep $t$. The summation extends over the $J$
internal energy  states $E_{int,j}$ of the $J$ distinct
geometric configurations
of the same cluster class $C(a_{f})$ (as defined in the Appendix).
Physically these energies simulate different excitation states of a
nuclear fragment of mass number $a_{j}$.  The factor $g_{j}$ is the
statistical weight of the energy state $E_{int,j}$.

To compute the
parameters referring to the internal states we construct the different
geometrically different cluster classes denoted
$G(a_{f},c_{f};\gamma)$ in the Appendix. We then evaluate the corresponding
energies, $E(a_{f},c_{f};\gamma)$, and the related multiplicities
$g(a_{f},c_{f};\gamma)$.  The full details are given in the Appendix.  The
factor $\lambda$ is the Lagrange multiplier that takes account of
conservation of nucleons.  The factor $h^{3}$ is the `volume' of an
elementary phase--space cell.  In the strict classical context of
statistical mechanics the elementary cell is not defined; a
quasi--classical argument identifies $h$ with
Planck's constant \cite{CH01}.  In the CA context we have a natural
phase space cell, inherited from the discretized space and discretized
velocity, or momentum space; the `volume' of this cell is $(\Delta
\lambda \times m \times v)^{3}$ which is to be substituted to $h^{3}$.

We re--write Eq. (15) in the form
$$ N(a_{f},t) \ = \
\Theta(T(t),V) \times
\sigma(a_{f}) \times
Z_{int}(a_{f},T(t)) \times
\exp (\lambda\  \nu_{f} \ a_{f}).\eqno  (15.a)$$
The first factor of the righthand--side, $\Theta(T(t),V)$
is essentially the translational partition function
of the clusters of class $C(a_{f})$
(normalised to $a_{f}$=1; the actual mass contribution,
${a_{f}}^{3/2}$, is included in the second factor
$\sigma(a_{f})$).
The translation contribution is evaluated
in the standard context of classical
mechanics, for reasons of algebraic simplicity
(the discrete kinetic energy states of the CA lead to a more
complicated expression, which should be equivalent to the
classical expression in the large $a_{f}$ limit).
This factor is the same for all species of clusters.
The second factor, $\sigma(a_{f})$ = ${a_{f}}^{1/2}/(a_{f}-1)!$,
is a combination of the mass effect
in the
kinetic energy contribution, and the effect of the
indiscernability of the nucleons
(invariance under permutation of all
nucleons of the cluster).
The third factor is essentially the internal partition function of
the cluster class $C(a_{f})$, which we evaluate in the specific
CA framework.

For the purposes of estimating reaction--equilibrium
temperatures from our numerical experiments we restrict
ourselves to clusters of smallest sizes, $a_{f}$ = 1, 2 and 3.
As transpires from the Appendix, the precise enumeration of the cluster
geometries becomes already rather involved for $a_{f}$ = 3.
To deal with higher sizes, we believe that
asymptotic approximations to the cluster configurations should be constructed.
This has not been done in the present work.

{\it $a_{f}$ = 1}

For a cluster made of a single nucleon we have $\sigma(1)$=1,
and the internal partition  function reduces to
\begin{equation}
Z_{int}(1,T(t)) \  =\ 1.
\label{eq:16}
\end{equation}

{\it $a_{f}$ = 2}

For a cluster made of 2 nucleons $\sigma(2)$=$\sqrt 2$;
the internal partition function involves
the contributions of the geometric
configurations listed under Eqs. A3--A5, and A6.
\begin{equation}
\begin{split}
Z_{int}(2,T(t)) \  = \
& \frac{\varphi}{2}\  \exp [-V_{\phi}/kT(t)] \
 +\ \frac{\epsilon}{2}\  \exp [-V_{e}/kT(t)] \\ 
& + \frac{\nu}{2}\  \exp [-V_{v}/kT(t)]
+\ 1 \ \exp [ -V_{o}/kT(t) ] \   .
\label{eq:17}
\end{split}
\end{equation}

{\it $a_{f}$ = 3}

A cluster of 3 nucleons has $\sigma(3)$=${\sqrt 3} / 2$.
The geometric configurations which contribute to
the partition function
are listed in the Appendix under Eqs. (A7)--(A9) (first line of Eq. (18)),
(A10)--(A15) (second and third lines),
(A16)--(A19) (fourth and fifth lines),
(A20)--(A22) (sixth line),
and (A23) (last line):
\begin{multline}
Z_{int}(3,T(t)) \  =
 \frac{\varphi}{2}\  \exp[-2V_{\phi}/kT(t)] \ +
\ \frac{\epsilon}{2}\  \exp [-2V_{e}/kT(t)] \ +
\ \frac{\nu}{2}\  \exp [-2V_{v}/kT(t)] \ + \\
(\varphi^{2}/2 - \varphi)\  \exp[-(2V_{\phi}+V_e)/kT(t)] \ +
\  (\epsilon^{2}/6 -\epsilon) \  \exp [-2V_{e}/kT(t)] \ +\\
\   \epsilon^{2}/6 \  \exp [-2V_{e}/kT(t)] \ +
\epsilon^{2}/6 \  \exp [-(2V_e+V_v)/kT(t)] \ +
\   \epsilon \  \exp [-2V_{v}/kT(t)] \ +\\
\   2\varphi \  \exp [-2V_{v}/kT(t)] \ + 
\frac{\varphi \epsilon}{3}  \exp [- (V_{\phi}+V_e)/kT(t)] \ +\\
\frac{\varphi \epsilon}{3}  \exp [-(V_{\phi}+V_e+V_v)/kT(t)] \ +
\frac{\varphi \nu}{2}  \exp [- (V_{\phi}+V_v)/kT(t)] \ +\\
\frac{3 \varphi\epsilon}{4}    \exp [-(V_e+V_v)/kT(t)] \ +
\varphi \  \exp [-(2V_{\phi}+V_o)/kT(t)] \ +\\
 \ \epsilon\  \exp [-(2V_e+V_o)/kT(t)] \ +
\nu\  \exp [-(2V_{v}+V_o)/kT(t)] \ +
1 \ \exp [ -3 (V_{o}+ \Delta V)/kT(t) ] \   .
\label{eq:18}
\end{multline}

The following ratios of cluster numbers are
independent of the Lagrange parameter $\lambda$:
$${ {N(3,t) N(1,t)} \over {N(2,t)^{2}} }\ =\
{{\sqrt 3} \over {4}} \times { {Z_{int}(3,T(t))} \over
{Z_{int}(2,T(t))^{2}} } , \eqno (19)$$
$${ {N(3,t)} \over {N(2,t) N(1,t)} }\ =\ {{\sqrt 3}\over{2 \sqrt 2 }}
\times
{ {1} \over {\Theta(T(t),V)} } \times
{ {Z_{int}(3,T(t))} \over {Z_{int}(2,T(t))} } , \eqno (20.a)$$
$${ {N(2,t)} \over {{N(1,t)}^{2}} }\ =\ {\sqrt 2} \times
{ {1} \over {\Theta(T(t),V)} } \times
 Z_{int}(2,T(t))   . \eqno (20.b)$$
The first ratio is independent of the translational term
$\Theta$ (and hence independent of the reaction volume $V$).
 The numbers of clusters $N(a_{f},t)$,
$a_{f}$=1,2,3 are directly supplied by our CA runs, at every timestep $t$, so
that the lefthand--sides of Eqs.  (19, 20.a, 20.b) are known.  These
equations are then solved with respect to the parameter $T(t)$
(written $T_{13/22}(t)$, $T_{3/12}(t)$, $T_{2/11}(t)$ respectively,
under the above alternatives).  These temperature parameters are the
reaction temperatures of the equilibrium processes
$$ C(1) \ + \ C(3)\ \rightleftharpoons\  2\ C(2)\ ,\ \
 C(3)\        \rightleftharpoons\  C(1)\ +\ C(2)\ ,\ \
 C(2)\        \rightleftharpoons\  2\ C(1)\ ,$$
respectively.

In order to follow as closely
as possible the actual laboratory experiment,
consider the cumulated number of
clusters of size $a_{f}$, $M(a_{f},t)$,  which have passed the
collection of detectors up to timestep $t$. The relevant
timestep is identified with the length of our run,
$t_{max}$ $\approx$ 70 steps.
With our detectors situated at a finite distance $d_{D}$
($\approx$ 50) from the reaction centre,
the first clusters arriving at the detectors
were emitted from the collision site $d_{D}-R_{C}$
timesteps prior to the arrival time ($R_{C}$,
radius of the initial merged $T+I$ configuration, $\approx$ 10 in our
set--up). Accordingly, in a run of 70 steps the first
fragments have time to be reflected on the boundary
of our CA universe.
The actual choice of $t_{max}$ secures that the
clusters cannot reach the
detectors after reflection;
thereby the clusters arriving close to the
detectors cannot undergo any interactions with other clusters.

 On the other hand, the last fragments registered by the detectors,
 at step $t_{max}$ have left the central nucleus roughly at step
 $t_{max} - (d_{D}-R_{C})$ ($\approx$ 30).
 Accordingly, the cumulated number readings of the detectors,
 terminating at timestep $t_{max}$ (=70), $M(a_{f},t_{max})$,
 cover the first 30 timesteps of the fragmentation mechanism.
 Alternatively, the cumulated number may be interpreted
 as the average number of clusters of size $a_{f}$ actually present
 in our system at the `average' time
 1/2 $[t_{max} - (d_{D}-R_{C})]$ $\equiv$ $t_{ref}$,
 of the order of 15:
 $$N(a_{f},t_{ref})\ \equiv\ M(a_{f},t_{max}). \eqno (21)$$
 In our numerical simulations we have set
 this reference time $t_{ref}$ equal to 17.

The three formal caloric curves $T_{13/22}(t)$--$E_{imp}/A$,
$T_{3/12}(t)$--$E_{imp}/A$, and $T_{2/11}(t)$ --$E_{imp}/A$,
for $t$ =  $t_{ref}$, are plotted in Fig. 5. All three curves are seen
to be essentially independent of the energy
over the range $E/A$ $<$25 MeV investigated:
$$T_{13/22}\ \approx
\  T_{3/21}\ \approx
\  T_{2/11}\ \approx\ 1\ - \ 2 \ {\rm MeV}.$$
Qualitatively this behaviour is in line with the different
formal temperatures derived from the laboratory Au--Au fragmentation as
analysed
in Trautmann \cite{NU97}
(with the exception of the He--Li isotope temperature).
Quantitatively the constant temperature level as found in the
laboratory experiments lies
at 5 Mev. The  experimental temperature is  thus shifted by $\Delta T$
$\approx$ 3.5--4 Mev with respect to the CA temperature:
$$ T^{Lab} \ = \ T^{CA} \ + \ \Delta T. \eqno (22)$$
In the case of the nucleon--gas temperature
we have noticed
a shift of a similar order between the theoretical CA temperature
and the laboratory estimate.

The temperature defect $\Delta T$ between the CA and experimental
temperature is thought to be due to a feature
inherent to our CA treatment. The construction of a stable nucleus in
the CA framework relies on a discretized version of classical
mechanics, in which the component nucleons possess no kinetic energy
at all (in the reference system attached with the centre of mass of
the nucleus).  Since classically temperatures are related to
microscopic kinetic energies, an underestimated kinetic energy
leads to underestimating the temperature as well.  The reaction
temperatures encoded in relations (19), (20.a), (20.b) refer to
small--size fragments only.  According to our remarks on the
cluster--gas temperature, these temperatures are not thought to be
representative for the real collection of fragments at high energies
($E_{imp}/A$ $>$ 8 MeV).  However, in the low energy domain the available
energy can concentrate on small--size fragments, which then form and
dissolve easily; therefore the latter fragment distribution does
reflect a physically meaningful temperature at lower energies.  The full
physically significant temperature run with energy (caloric curve),
from 0 to about 25 MeV, is suggested therefore to be made of the
cluster temperature $T_{c(C)}$ at the high energy end, and the
reaction temperatures ($T_{2/11}$, etc) at the low energy end, with a
continuous transition from one curve to the other around the
critical energy $E_{c}/A$ $\approx$ 7--8 MeV/nucleon,
of the order of the average binding energy.

\section{ Conclusion}

The primary aim of the proposed CA simulation has been to devise a framework
capable of replicating the actual laboratory procedure of monitoring a
fragmentation process generated in cluster collisions.  Previous theoretical
work, investigating thermodynamic properties of the collection of
fragments (e.g., a caloric curve), was implicitly based on an
assumption of a thermodynamic equilibrium.  In the present model no
equilibrium hypothesis is needed.  Statistical
relations (Eqs 10, 11; 13; 19, 20.a,b) are used formally, for the
purpose of making comparisons with the laboratory experiments.

Our numerical experiments demonstrate
that the distribution of the CA clusters against cluster size,
as registered by a collection of detectors surrounding
the collision site, obeys a power law of exponent $\tau$ close to 2.6.
This model result is in excellent agreement with the laboratory results
of nuclear as well as other multifragmentation processes.
It is well
known that the very existence of a power law can be derived in the context
of various statistical models (cf the review \cite{RI01}).
The statistical
assumptions of these theoretical approaches, and the related counting
procedure, do not respect, however, the real laboratory protocol.  It
should then not come as a surprise that the theoretically derived
slope of the
(equilibrium) distribution is found to be significantly different from
the experimentally measured slope of the (far from equilibrium)
distribution (2.3 against thexperimental value 2.6).

Secondly, the CA experiments are indicative that the notion of
a compound cluster, which would support a meaningful thermodynamic
treatment, is of limited value in the fragmentation problem.  We
observe that typically during the earliest phases of the collision
process the combined $T+I$ is fractured into a few large fragments.
The latter tend to acquire all the available mass when the impact
energy becomes large enough; in the latter limit, $E_{imp}$ $\gg$
binding energy, two fragments survive.  The target $T$ then becomes
transparent to the incident cluster $I$.  Significant energy sharing
is found to occur in the range of low impact energies.

Thirdly, our CA model experiments demonstrate that the original
spatial symmetry of the collision set--up is statistically
preserved during the whole fragmentation mechanism (Fig. 2).

Finally, a major goal of our Paper was to construct
formal  caloric curves,  $T^{CA}$ -- $E_{imp}/A$,
based on a variety of formal temperature parameters
directly derived from the CA experiments, and defined and discussed in this
Paper for clusters of particles of arbitrary nature.  In the nuclear
fragmentation set--up, comparison of our CA caloric curves (temperature
parameter chosen at a selected timestep $t_{ref}$), with the
laboratory caloric curves, $T^{Lab}$--$E_{imp}$, derived from the Au--Au
collision experiments in the case of different methods of measurement
of a temperature $T^{Lab}$, calls for several comments.

With none of the formal temperatures introduced
we can reproduce the qualitative shape of the full NuPECC
caloric curve \cite{NU97}. The theoretical treatment does not reveal
a transition from a rising behaviour to a plateau,
and again from a plateau to a rising branch.
Full CA caloric curves either reduce to a plateau (case of the
reaction temperatures, Eqs.  (19), (20.a), (23.b)); or else the curves
are rising everywhere (nucleon gas, Eqs.  (10), (11); cluster gas, Eq.
(13)).
The formal CA temperatures
are qualitatively more in line with the Trautmann analysis
of the experimental Au--Au.
Experimental reaction temperatures are found to be
independent of the excitation energy per nucleon;
this behaviour is duplicated for our 3 CA reaction temperatures.
The laboratory He--Li temperature parameter exhibits a rising
behaviour, reminiscent of the rising pattern of the
temperatures of the
nucleon gas and the cluster gas.
The slope of the latter experimental curve, of approximately  0.27
(in the energy range 0 to 15  MeV), compares favourably with
the average slope of the cluster--gas, of 0.3,
in the energy range $<$ 7.5 MeV.
Quantitatively the CA temperatures are typically too low
as compared to the laboratory measurements of the temperature parameters, by an
amount $\Delta T$ of nearly 4 MeV (Eq.  22).  We have traced this
effect to the treatment of the dynamics of the nucleons in a context
of classical mechanics; in this framework the residual quantum
mechanical zero--point kinetic energy is disregarded.  We believe that
this effect may account for a defect in the theoretical temperatures.
An extension of the model taking account of quantum--mechanical
effects will eventually be needed to handle the energy problem
adequately (cf \cite{KO93} for an attempt at implementing quantum
mechanics in the CA framework).

In the specific nuclear multifragmentation case we have analysed, besides
the absence of quantum corrections, our model ignores any
charge--related effects, so that the detailed theoretical results
cannot be compared directly with the detailed experimental
measurements.  All laboratory cluster identifications rely on counts
of isotopes (rather than isobars, as done in our approach).  We hope
to be able to include electrostatic effects at a later stage.  We also
plan to extend the model to handle asymmetric collisions ($A_{I} \ne
A_{T}$) and collisions involving a non--zero impact parameter.

The CA model as developed
in this Paper remains a manifestly highly schematic representation
of an actual
physical fragmentation problem of any nature, and as such it can only
replicate, and hence also isolate,  properties which are largely
insensitive to the microscopic details of the physics.  Our numerical
experiments suggest that the power law of the fragments, and on a
quantitative level, the exponent of the latter, belong into this
category of invariant properties.

\vskip 1cm
\section { Acknowledgments}

A.L. would like to thank the SPM Department of the CNRS (Paris)
 for financial support for a stay
at LPT in Strasbourg during which this work was initiated.
He thanks the members of LPT for the hospitality extended to him.
J.P. gratefully acknowledges  several
Royal Society--FNRS European Exchange
Fellowships (1999-2001). He wishes to thank the Fellows  of
New Hall College, Cambridge, UK,  for their kind hospitality.

\newpage


\section { Appendix : Enumeration of Cluster Configurations}

\subsection{ Clusters. Cluster equivalence classes. Cluster geometries}

We introduce a definition of
a `cluster' of size $a$ that rests on the notion of
`interaction neighbourhood' $N_{\rm int}({\bf r}) $ \cite{AL99}.
If ${\bf r}$ labels an arbitrary
cell, then any cell ${\bf r'}$ distinct from $\bf r$ and contained in
the interaction neighbourhood of  cell $\bf r$, $N_{\rm int}({\bf r})$, is
termed
`adjacent' to cell ${\bf r}$.  The notion of adjacency is naturally
extended to an arbitrary set of cells, $S$: A cell $\bf r' $ is
adjacent to the set of cells $S$ if (i) $\bf r' $ $\notin$ $S$; and
(ii) there is a cell ${\bf r}$ $\in$ $S$ such that $\bf r'$ $\in$
$N_{\rm int}(\bf r)$.  The collection of all cells adjacent to $S$ is
the outer boundary of $S$, $\partial S$.

Define a `walk' in the CA, of head ${\bf r}_{o}$ and tail ${\bf r}_{f}$,
$w({\bf r}_{o}, {\bf r}_{f})$,
as an ordered collection of cells
$({\bf r}_{o}, {\bf r}_{1}, \ldots, {\bf r}_{f-1}, {\bf r}_{f})$,
such that for any pair of successive cells  $({\bf r}_{j}, {\bf r}_{j+1})$
we have
${\bf r}_{j+1})$ $\in$ $N_{\rm int}({\bf r}_{j})$
(cf [24] for the graph--theoretical details).
Define further a `connected set of cells', $Q$,
in the lattice space of the CA as a set of cells
$ \{ {\bf r}_{a},{\bf r}_{b},\ldots \} $,
such that for any pair of cells of the set,
${\bf r}_{n}, {\bf r}_{m}$ $\in$ $Q$, there exists a walk
$w({\bf r}_{n}, {\bf r}_{m})$ $\in$ $Q$
(all cells of the walk lie in the connected set).

We understand by a `cluster rooted at particle ${\alpha}$', $F[{\alpha}]$,
the connected set of cells such that
(i) particle ${\alpha}$ is located in one  cell of the set;
(ii) each cell of the set is non--empty (it contains at least one  particle);
and (iii) the outer boundary of the connected set, $\partial F[{\alpha}]$,
is empty.

A cluster rooted at particle $\beta$,  $F[{\beta}]$,
is identical with the cluster rooted at particle $\alpha$, $F[{\alpha}]$,
$F[{\beta}]$ $\equiv$ $F[{\alpha}]$,
if particle $\beta$ occupies a cell of the connected set of cells
$F[{\alpha}]$.
Accordingly we can talk about `cluster  $F$', without
reference to a root particle.
A `cluster', or `fragmentation cluster',  $F$,
is a  connected set of non--empty cells,
which has an empty boundary.

We distinguish the fragmentation clusters
occurring in a CA experiment by an argument $f$, writing
 $F(f)$  to refer to the  $f^{\rm th}$ `fragmentation cluster'
 in a given CA context (at a given timestep $t$).
Any collection of clusters $F(i)$, $F(j)$, $\ldots$,
containing the same number of particles,
regarded as indiscernible,
$a_{i}$ = $a_{j}$ = $\ldots$ $\equiv$  $a_{f}$, belongs into
the same `cluster equivalence class' $C(a_{f})$.
It is the cluster equivalence class
$C(a_{f})$ which we regard as
the CA counterpart of the `nuclear fragment' of mass number
$A$ = $a_{f}$ in the nuclear fragmentation laboratory experiment.

A pair of clusters $F(i)$, $F(j)$,
of same cluster equivalence class,
in which the cells are joined according to a same geometric rule,
and such that each cluster of the collection has same binding energy,
will be said  to have same `cluster geometry' $G$.
The collection of all clusters (of same cluster equivalence class)
which have same cluster geometry define the
`cluster--geometric equivalence class' $G$.

Two clusters $F(i)$, $F(j)$ of same cluster geometry $G$
are said to be `geometrically equal'
if after translation along the lattice axes,
and permutation of the particles among the cells,
they can be superposed exactly.
Otherwise the two configurations are geometrically unequal or distinct.
The number $g$ of geometrically distinct configurations
of a cluster--geometric equivalence class $G$
is the `multiplicity' of the
cluster geometry.

The reason for assigning a special status to clusters superposable
under translations is that the statistical treatment of Section 5
deals separately with the translational motions of the fragments,
the effect on the reaction equilibrium of the energy attached to
these motions being  described by the function $\Theta(T,V)$ (Eq. 15.a)
(a translational partition function).
The evaluation of the internal partition function $Z_{int}$,
which concerns us here, involves those configurations
which we have referred to as `geometrically distinct'.

\subsection{ Construction of cluster geometries}

Essentially, the notion of cluster geometry
enables us to group together clusters
which are geometrically distinct, but which
become superposable after application of certain
geometric transformation groups
(the discrete lattice symmetries).
It is manifest that clusters of same cluster--geometry
have generically same binding energies.
Conversely, if the interaction energies are generic,
then different cluster geometries realize different
binding energies (accidental degeneracies may occur as a result
of a non--generic interaction potential).
For the purposes of the statistical mechanics of Section 5
`energy equivalence classes' rather than cluster--geometric
equivalence classes have to be isolated. The above comment
indicates that the two questions are
essentially equivalent (or closely related in the case of
degeneracies).

We have the following natural construction procedure for a
cluster geometry, $G(a,c;\gamma)$, which we specify
(i) by the number of particles $a$ in the cluster;
(ii) the number of cells $c$ of the cluster;
(iii) the precise mode of assembling the cells,
and the precise distribution of the $a$ particles among the $c$ cells,
which we symbolize by the descriptive parameter $\gamma$:

(1) Assemble the $c$ different cells according to the
selected construction rule $\gamma$ to form a connected set.

(2) Distribute the $a$ ($\geq c$)
particles among the $c$ cells; this procedure generates one representative
of the cluster--geometry $G(a,c;\gamma)$.

(3) To find the multiplicity $g(a,c;\gamma)$
of the cluster geometry, generate
all geometrically distinct  configurations
representative of the same cluster geometry;
this is done by applying  the symmetry operations of the lattice
(barring translations along the axes,
as well as those geometric symmetries which are equivalent to particle
permutations).
The binding energy of a representative of this cluster geometry,
naturally denoted by  $E(a,c;\gamma)$,
is obtained by applying the rules of Section 2.

The totality of cluster--geometries corresponding to a
cluster $C(a)$ is finally generated by repeating steps (1), (2)
and (3) for all allowed  choices of cells,
$c$ = $a$, $a-1$, $\ldots$, 1,
and all possible geometrical assemblages of the cells into clusters.

For $a_{f}$ = 1 there is only one cluster equivalence class;
the binding energy is zero.
For any $a_{f}$ $>$ 1 the cluster equivalence class
of a specied cluster $C(a_{f})$ contains several cluster geometries
 $G(a_{f},c_{f};\gamma)$, which are energetically distinct.

In order to carry out the construction and characterization
of the cluster geometries it is helpful to resort to
an algebraic notation for the parameter $\gamma$:

(1) If $c_{f}$ = 1 no construction is involved; we then write
$\gamma$  $\equiv$  $\emptyset$.

(2) If  $c_{f}$ = 2 there are 3 distinct modes of contact of the two
cubic cells, generated by
face--wise, edge--wise, or  vertex--wise joining of the cells;
these modes specify  the cluster geometry completely.
We write
$\gamma$ $\equiv$ $\phi$ (face--joining); or
$\gamma$ $\equiv$ $e$ (edge--joining); or
$\gamma$ $\equiv$ $v$ (vertex--joining) respectively.

Since for instance face--joining $\phi$ is not a geometrically
unambiguously defined operation in the CA lattice space,
we introduce more elementary construction symbols which specify
unique and independent operations;
the elementary (and in this specific case irreducible)
independent operations
are the joinings  along the lattice axes: $x$--axis, $\phi_{x}$;
$y$--axis, $\phi_{y}$;
and  $z$--axis, $\phi_{z}$.
(We can join the second cell to the first cell in the positive
or in the negative $x$--direction; however, the two resulting
configurations are
superposable under translations along the $x$--axis; they count as
`geometrically equal').

These construction symbols can be combined by a first operation,
of addition (+)
$$\phi_{x}\  + \ \phi_{y}\  + \ \phi_{z}  = \phi ;$$
the sign (+) is read `or': face--joining is face joining
along the $x$--axis,
{\it or} along the $y$--axis,
{\it or} along the $z$--axis,
and this exhausts the possible alternatives.
Denote
the number of cell--faces in the $x$--direction by
$\varphi_{x}$, etc; we then have
$\varphi_{x}$ (= $\varphi_{y}$ =$\varphi_{z}$) = $\varphi/3$ (=2).
The multiplicity of the geometric configuration generated by $\phi_{x}$ is
1 (=$\varphi_{x}$/2, the division by 2 being due to the geometrical
equivalence of the joining along the `positive' or `negative' face).
The multiplicity of the geometric--cluster class $G(2,2;\phi)$ is
then the sum of the multiplicities
of the component elementary geometric--cluster classes
$G(2,2;\phi_{x})$, etc.
This property of additivity of multiplicities is an instance of the
following obvious property:

If a general construction procedure $\gamma$ is the sum (+)
of independent elementary construction procedures,
$\gamma_{1}$, $\gamma_{2}$, $\ldots$,
then the multiplicity of the cluster--geometry is the
sum of the multiplicities of the corresponding elementary cluster--geometries:
$$\gamma\ = \ \gamma_{1}\ + \ \gamma_{2}\  + \ \ldots \ + \ \gamma_{n}: \ \
   g(a,c;\gamma) \ = \ g(a,c;\gamma_{1}) \ + \ g(a,c;\gamma_{2}) \ +
\ldots \  + \ g(a,c;\gamma_{n})\ . \eqno (A1)$$
A similar break--up of the operation of edge--joining $e$ holds,
 $e$ = $e_{x}$ +$e_{y}$ +$e_{z}$;
where $e_{x}$ is the more elementary (though not
irreducible) operation of joining two cells along an edge
parallel to the $x$--axis. We have, with obvious
notations,
 $\epsilon_{x}$ (= $\epsilon_{y}$ = $\epsilon_{z}$) = $\epsilon/3$ (=4)
different edges parallel to the $x$--axis, and, as in the case of
face--joining, half of these edges (=$e_{x}$/2) produce
geometrically distinct configurations (the joining along opposite edges
with respect to the centre of the cube generates geometrically
equal configurations; the joining along edges belonging to a same
face produces geometrically distinct configurations).

Similar considerations hold for vertex--joining.

(3) If  $c_{f}$ = 3, first form a 2--cell cluster,
joining two cells as under (2);
the third cell is then attached to the 2--cell cluster.
A sequence of two joining operations is
indicated algebraically by a multiplication sign (.) between these
operations.
For instance, the combined operation
$e_{y}.e_{x}$
symbolizes the  construction of
3--cell clusters, a pair of cells being joined along an $x$--edge; the
attachment of the third cell is along a $y$--edge.
Through addition and multiplication of elementary construction operations
more complex
construction schemes can be generated.

For $c_{f}$ = 3 we have 6 main combinations of joinings,
($\phi\ .\ \phi$, $e\ .\ e$, $v\ .\ v$,
$\phi\ .\ e$, $e\ .\ v$, $v\ .\ \phi$); each main combination
gives rise to directional variants which are described in
terms of the elementary operations ($\phi_{x}$, etc).
As an instance, take the  face--wise joining of all 3 cells,
the common faces being  parallel; this operation is symbolised by
$\phi_{x}\ .\ \phi_{x}$ + $\phi_{y}\ .\ \phi_{y}$ + $\phi_{z}\ .\ \phi_{z}$.

In the construction of cluster geometries  we have to take account
very often of special clauses in the combination of the elementary
operations. For instance, among the different
edge--joinings of 3 cells,
consider the joining along
$x$--edges combined with the joining along $y$--edges,
($e_{y}\ .\ e_{x}$), under the extra constraint that
both edges have no vertex in common.
Such extra clauses are symbolised
by appropriate subscripts to the construction symbol.
In our illustration the absence of a common vertex is indicated
by the subscript $-v$, so that the construction symbol
becomes $(e_{y} . e_{x})_{-v}$. In a similar fashion
the symbol $(e_{y} . e_{x})_{+v}$ points out that the
two edges are required to have a common vertex (subsript $+v$).
We have
$$e_{y} . e_{x} \
= \ (e_{y} . e_{x})_{-v}\ + \ (e_{y} . e_{x})_{+v}.$$
(This notation is consistent  with the
symbols for the elementary operations: in $\phi_{x}$
the subscript $x$ indicates a constraint to the general face--joining
operation  $\phi$).

The trivial cluster equivalence class $C(1)$ contains a single
cluster geometry, $G(1,1;\emptyset)$ (single non-empty cell).
The corresponding binding energy is  zero,
$$G(1,1;\emptyset)\ : \
\ E(1,1;\emptyset) \ = \ 0 \ , \  g(1,1;\emptyset) \ = \  1. \eqno (A2)$$

\subsection{ Cluster equivalence class $C(2)$ }

The cluster equivalence class $C(2)$ is (a) either a collection of $c$ = 2
adjacent cells containing one particle each; (b) or it is a single cell,
$c$ =1, containing both particles.

(a) Under the first alternative we have all three modes of joining the
two cells, $G(2,2;j)$, $j$ = $\phi$, $e$, or $v$:

(i) Common face: geometry $G(2,2;\phi)$.

Multiplicity :
$$g(2,2;\phi)\  =\
g(2,2;\phi_{x})\ +\  g(2,2;\phi_{y})\  + \ g(2,2;\phi_{y}). $$
Since $g(2,2;\phi_{x})$ = $\varphi_{x}$/2 (cf above),  we have
$$G(2,2;\phi) \ :\ \ E(2,2;\phi) \ = \ V_{\phi}\ ,
\ g(2,2;\phi) \ = \ {{\varphi} \over {2}} \ =\ 3. \eqno (A3)$$

(ii) Common edge: geometry  $G(2,2;e)$.

Proceeding as under (i) (with substitution
$\phi$ $\to$ $e$, and $\varphi$ $\to$ $\epsilon$), we have
$g(2,2;e_{x})$ = $\epsilon_{x}$/2, and hence
$$G(2,2;e)\ :\ \ E(2,2;e) \ = \ V_{e}\ ,
\ g(2,2;e) \ = \ {{\epsilon} \over {2}} \ = \ 6. \eqno (A4)$$

(iii) Common vertex: geometry  $G(2,2;v)$.

In (i) substitute $\phi$ $\to$ $v$,
and $\varphi$ $\to$ $\nu$, and consider
vertices lying on the main diagonals of the cubic cell; take acount that
the joining in opposite
directions along a diagonal generates geometrically equal configurations.
Therefore
$$G(2,2;v)\ :\ \ E(2,2;v) \ = \ V_{v}\ ,
\ g(2,2;v) \ = \ {{\nu} \over {2}} \ = \ 4. \eqno (A5)$$

(b) The second alternative is trivial
$$G(2,1;\emptyset)\ :\ \  E(2,1;\emptyset) \ = \ V_{o}, \
g(2,1;\emptyset) \ =\ 1. \eqno (A6)$$

\subsection{ Cluster equivalence class $C(3)$ }

This class $C(3)$ gives rise to 3 broad
categories of geometrical clusters,
(a) $G(3,3;\gamma)$,
(b) $G(3,2;\gamma)$,
and (c)  $G(3,1;\emptyset)$.

(a) Category $G(3,3;\gamma)$ contains a first variety
of 3 geometries of type $G(3,3;j\ .\ j)$, obtained by attaching one cell
to each of the 3 geometries
$G(2,2;j)$, ($j$ designates an operation
$\phi$, or $e$, or $v$, as under $C(2)$; if the first joining is
a face--joining, then so is the second; etc); this variety of
geometrical clusters is constrained to form three aligned cells.

(i) Pair of common opposite faces: geometry $G(3,3;(\phi\ .  \phi)_{-e})$
(three cells aligned along a lattice axis); subscript
$-e$ indicates that the two faces of the cubic cell have no common
edge; hence they are parallel.

The construction procedure is explicited as follows
$(\phi\ .  \phi)_{-e}$ =
 $\phi_{x} \ .\ \phi_{x}$
+ $\phi_{y}\ .\ \phi_{y}$
+ $\phi_{z}\ .\ \phi_{z}$;
hence the multiplicity from Eq (A1):
$$G(3,3;(\phi\ .\  \phi)_{-e})\ :
\ E(3,3;(\phi\ .\  \phi)_{-e}) \ = \ 2 V_{\phi}\ ,\
\ g(3,3;(\phi\ .\  \phi)_{-e}) \ = \ {{\varphi} \over {2}} \ =\ 3. \eqno (A7)$$

(ii) Pair of diametrically opposite  common edges:
geometry  $G(3,3;(e\ . \ e)_{-\phi})$ (three cells aligned along diagonals
of one face); subscript
$-\phi$ indicates that the two edges of the cubic cell do not belong
to a common face; this implies that
they are symmetric with respect to the cell centre.
$$G(3,3;(e\ . \ e)_{-\phi})\ : \
\ E(3,3;(e\ . \ e)_{-\phi}) \ = \ 2 V_{e}\ ,\
\ g(3,3;(e\ . \ e)_{-\phi}) \ = \ {{\epsilon} \over {2}} \ =\ 6. \eqno (A8)$$

(iii) Pair of diametrically opposite common vertices:
geometry  $G(3,3;(v\ . \ v)_{-\phi})$ (three cells aligned along cell
diagonal); as under (ii) subscript
$-\phi$ stresses that the two vertices do not belong
to a common face; hence they are symmetric with respect to the cell centre.
$$G(3,3;(v\ . \ v)_{-\phi})\ : \
\ E(3,3;(v\ . \ v)_{-\phi})\ = \ 2 V_{v}\ ,\
\ g(3,3;(v\ . \ v)_{-\phi}) \ = \ {{\nu} \over {2}} \ =\ 4. \eqno (A9)$$

Category $G(3,3;\gamma)$ includes a second variety
of 3 geometries of type $G(3,3;j\ .\ j)$, this time
under the extra constraint of non--alignment of the  three cells.
(The two constraints, alignment, and non--alignment of the three
cells, joined by a same operation $j$,
manifestly exhaust all alternatives of type $G(3,3;j\ .\ j)$).

(i$'$) Pair of non--parallel common faces:
geometry $G(3,3;(\phi\ . \  \phi)_{+e})$
(L--shaped configuration); subscript $+e$ points out that
the two faces have a common edge.

Start with a central cube,
and add a pair of cubes having faces in common with the central cube;
there are $\varphi(\varphi-1)/2$ different pairs, among which
$\varphi/2$ pairs belong to alternative (i) already listed.
All of these configurations
have same energy.
$$G(3,3;(\phi\ . \  \phi)_{+e})\ :\
\ E(3,3;(\phi\ . \  \phi)_{+e}) \ = \ 2 V_{\phi}\ +\ V_{e} \ ,\
\ g(3,3;(\phi\ . \  \phi)_{+e}) \ = \ { {\varphi}^{2}  \over
{2} }  \ - \varphi =\ 12. \eqno (A10)$$

(ii$'$) The case of non--diametrically opposite common edges
gives rise to 3 geometry classes.

($\alpha$) Parallel joining edges: The construction scheme is
$$(e\ .\ e)_{+\phi} \ =\
 (e_{x}\ .\ e_{x})_{+\phi}\ + \ (e_{y}\ .\ e_{y})_{+\phi}\  + \
 (e_{z}\ .\ e_{z})_{+\phi}, $$
where the subscript $+\phi$ signifies that the two parallel edges
belong to a same face of the cube.
The multiplicity of $(e_{x} . e_{x})_{+\phi}$ is  (cf  i$'$)
${\epsilon_{x}}^{2}/2 -  \epsilon_{x}$,
with $\epsilon_{x} = \epsilon/3$. From Eq (A1) we obtain the total
multiplicity.
$$G(3,3;(e\ .\ e)_{+\phi})\ :\
\ E(3,3;(e\ .\ e)_{+\phi}) \ = \ 2 V_{e} \ ,\
\ g(3,3;(e\ .\ e)_{+\phi})\ =
\   { {\epsilon}^{2}  \over {6} }\  - \epsilon\  =\ 12. \eqno (A11)$$

($\beta$) The two joining edges are distinct, $e$, $e'$
(mutually  orthogonal), and have
no common vertex (subscript $-v$); (alternatively we may say that
the two edges do not belong to the same face). Hence the construction
scheme is
$$(e\ .\ e')_{-v}\ =\
(e_{x}\ .\ e_{y})_{-v}\ +\ (e_{y}\ .\ e_{z})_{-v}\ + \ (e_{z}\ .\
e_{x})_{-v}. $$
The multiplicity of $(e_{x}\ .\ e_{y})_{-v}$ is (cf  i$'$)
$ \epsilon_{x}\epsilon_{y}/2$.
Hence
$$G(3,3;(e\ .\ e')_{-v})\ :\
\ E(3,3;(e\ .\ e')_{-v}) \ = \ 2 V_{e} \ ,\
\ g(3,3;(e\ .\ e')_{-v}) \ = \ { {\epsilon}^{2}  \over
{6} }\   =\ 24. \eqno (A12)$$
($\gamma$) The two joining edges are distinct
(mutually orthogonal), and do have
a common vertex (subscript $+v$) (or they do belong to a same face).
Under this alternative  each cell shares an edge
with each of its two neighbour cells.
The construction scheme is
$$(e\ .\ e')_{+v}\ =\
(e_{x}\ .\ e_{y})_{+v}\ +\ (e_{y}\ .\ e_{z})_{+v}\ + \ (e_{z}\ .\
e_{x})_{+v}. $$
The multiplicity of the geometry generated by $(e_{x}\ .\ e_{y})_{+v}$
is as under ($\beta$), namely
$ \epsilon_{x}\epsilon_{y}/2$.
$$G(3,3;(e\ .\ e')_{+v})\ :\
\ E(3,3;(e\ .\ e')_{+v}) \ = \ 2 V_{e}\ +\ V_{v}\ ,\
\ g(3,3;(e\ .\ e')_{+v}) \ =
\ { {\epsilon}^{2}  \over {6} }\   =\ 24. \eqno (A13)$$
(iii$'$) Pair of non--diametrically opposite common vertices.
These configurations fall into 2 geometry classes.

($\alpha$) Two vertices lying on same edge: geometry $G(3,3;(v\ .\ v)_{+e})$
(subscript $+e$ indicating that the vertices are joined by a common
edge).
We decompose this construction scheme as follows
$$(v\ .\ v)_{+e} \ =\
\ (v\ .\ v)_{+e_{x}}\ +
(v\ .\ v)_{+e_{y}}\ +
(v\ .\ v)_{+e_{z}},$$
with the notation $(v\ .\ v)_{+e_{x}}$ indicating that the two
vertices are constrained to lie on an $e_{x}$ edge, etc.
The multiplicity of $(v\ .\ v)_{+e_{x}}$ is
 $\epsilon_{x}$ = 3.
$$G(3,3;(v\ .\ v)_{+e})\ :\
\ E(3,3;(v\ .\ v)_{+e}) \ = \ 2 V_{v} \ ,\
\ g(3,3;(v\ .\ v)_{+e}) \ =
\    \epsilon =\  12. \eqno (A14)$$
($\beta$) Two vertices belonging to different edges of same face
(vertices belonging to different faces are diametrically opposite
and have been dealt with under (iii) A9). It is convenient
to describe this alternatively by saying
that these vertices belong to the same  diagonal $\delta$
of a face (constraint indicated by subscript $+\delta$):
$G(3,3;(v\ .\ v)_{+\delta})$.
If we denote by $\delta_{x}$ the collection
of the diagonals of the faces normal to
the $x$--lattice direction, etc, the relevant
construction scheme becomes
$$(v\ .\ v)_{+\delta} \ =
\ (v\ .\ v)_{+\delta_{x}}\ +
(v\ .\ v)_{+\delta_{y}}\ +
(v\ .\ v)_{+\delta_{z}}.$$
Since the multiplicity of $(v\ .\ v)_{+\delta_{x}}$
is readily seen to  be  $2 \varphi_{x}$ = 4, we have
$$G(3,3;(v\ .\ v)_{+\delta})\ : \
\ E(3,3;(v\ .\ v)_{+\delta}) \ = \ 2 V_{v} \ ,\
\ g(3,3;(v\ .\ v)_{+\delta}) \ =
\    2 \varphi =\  12. \eqno (A15)$$

Finally, category $G(3,3;\gamma)$ generates a third variety
of 3 geometries of type $G(3,3;j\ .\ j')$, where $j$ and  $j'$
are of different
nature (if $j$ stands for face--joining, $j'$ must be
either  edge--joining or vertex--joining, etc).

(iv) Face and edge--joining:
geometries $G(3,3;e\ .\ \phi)$.
These configurations fall into 2 geometry classes.

($\alpha$) Edge--joining on `small side' of the face--joined box:
$G(3,3; e_{\parallel \phi}\ .\ \phi)$ (where  $e_{\parallel \phi}$
indicates that the edge is parallel to the face $\phi$).
Construction scheme:
$$e_{\parallel \phi}\ .\ \phi \ = \
e_{y}\ .\ \phi_{x} \ + e_{z}\ .\ \phi_{x} \ +
e_{z}\ .\ \phi_{y} \ + e_{x}\ .\ \phi_{y} \ +
e_{x}\ .\ \phi_{z} \ + e_{y}\ .\ \phi_{z} \ .$$
The multiplicity of the geometries resulting from
$e_{y}\ .\ \phi_{x}$ is  $\varphi_{x}\epsilon_{x}/2$ (=4), hence
$$G(3,3; e_{\parallel \phi}\ .\ \phi)\ :\
\ E(3,3; e_{\parallel \phi}\ .\ \phi) \ = \ V_{\phi}\ +\ V_{e}\ ,\
\ g(3,3; e_{\parallel \phi}\ .\ \phi) \ =
\ {{\varphi \epsilon} \over {3}}\  =\ 24. \eqno (A16)$$

($\beta$) Edge--joining on `long side' of the face--joined box:
$G(3,3; e_{\perp \phi}\ .\ \phi)$ (where  $e_{\perp \phi}$
indicates that the edge is normal to the face $\phi$).
Construction scheme:
$$e_{\perp \phi}\ .\ \phi \ = \
e_{x}\ .\ \phi_{x} \  +
e_{y}\ .\ \phi_{y} \  +
e_{z}\ .\ \phi_{z} \  .$$
Since the number of edges of the `long side' is 2$\epsilon_{x}$,
the multiplicity of $\phi_{x}e_{x}$ is  $\varphi_{x}\epsilon_{x}$ (=4).
$$G(3,3; e_{\perp \phi}\ .\ \phi)\ :\
\ E(3,3; e_{\perp \phi}\ .\ \phi) \ = \ V_{\phi}\ +\ V_{e}\ + V_{v}\ ,\
\ g(3,3; e_{\perp \phi}\ .\ \phi) \ =
\ {{\varphi \epsilon} \over {3}}\  =\ 24. \eqno (A17)$$

(v) Face and vertex--joining (third cell attached to an outer free
vertex of box):
geometries $G(3,3;v\ .\ \phi)$.
Construction scheme
$$v\ .\ \phi \ = \ v\ .\ \phi_{x}\  +\ v\ .\ \phi_{y}\ +\ v\ .\ \phi_{z} . $$
The multiplicity of $v\ .\ \phi_{x}$ is $\nu \varphi_{x}/2$
$$G(3,3;v\ .\ \phi)\ :\
\ E(3,3;v\ .\ \phi)\ = \ V_{\phi}\ + \ V_{v}\ ,\
\ g(3,3;v\ .\ \phi)\ =
\ {{\varphi \nu} \over {2}}\  =\ 24. \eqno (A18)$$

(vi) Edge and vertex--joining (two cells joined vertex--wise; third
cell attached to an edge not incident to the joining vertex):
geometries $G(3,3;v\ .\ e)$.
Construction scheme: Attach second cell to
a vertex of the first cell ($\nu/2$ distinct alternatives, according to
$G(2,2;v)$);
the third cell is to be added such as to have a common vertex (but not
a common face) with the two cells; we have $3/4 \epsilon$ (=9)
alternatives on each of the two cells. Hence the multiplicity
$$G(3,3;v\ .\ e)\ :\
\ E(3,3;v\ .\ e)\ = \ V_{e}\ + \ V_{v}\ ,\
\ g(3,3;v\ .\ e)\ =
\ {{3\ \varphi\epsilon} \over {4}}  =\ 72. \eqno (A19)$$

(b)  Category $G(3,2;\gamma)$ generates essentially
the variety  of geometries $G(2,2;\gamma)$, with one difference.
In the $G(2,2;\gamma)$ case the two cells are equivalent and
indistinguishable; in the  $G(3,2;\gamma)$ case, one cell contains 2
particles, while the other contains only one particle, so that
the translational symmetries do not hold. The multiplicities
are therefore twice the multiplicities encountered under $G(2,2;\gamma)$.

(i) Common face: geometry $G(3,2;\phi)$.
$$G(3,2;\phi)\ :\
\ E(3,2;\phi)\ = \ 2 V_{\phi} \ +\ V_o\ ,\
\ g(3,2;\phi)\ = \  \varphi  =\ 6. \eqno (A20)$$

(ii) Common edge: geometry  $G(3,2;e)$.

$$G(3,2;e)\ :\
\ E(3,2;e)\ = \ 2 V_e\ + \ V_o\ ,\
\ g(3,2;e)\ = \  \epsilon \ = \ 12. \eqno (A21)$$

(iii) Common vertex: geometry  $G(3,2;v)$.

In (i) substitute $\phi$ $\to$ $v$, and consider
vertices lying on the main diagonals; take acount that
the joining in opposite
directions along a diagonal produces an equivalent configuration
under combinations of translations.
Therefore
$$G(3,2;v)\ :\
\ E(3,2;v)\ = \ 2 V_{v}\ + \ V_{o}\ ,\
\ g(3,2;v)\ = \  \nu  \ = \ 8. \eqno (A22)$$

(c)  The final alternative of all 3 particles
in the same cell generates the trivial geometry
$$G(3,1;\emptyset)\ : \
\ E(3,1;\emptyset)\ = \ 3 V_{o}\ + \ 3\Delta V, \
\ g(3,1;\emptyset)\ =\ 1. \eqno (A23)$$

This completes our list of distinct  geometrical configurations
associated with the cluster classes $C(1)$, $C(2)$ and $C(3)$.

The systematic construction method of the configurations extends
naturally to clusters of arbitrary size $a$.
However, with increasing $a$
the number of alternatives increases exponentially,
so that a detailed enumeration becomes rapidly prohibitive.
This suggests that an asymptotic approach, with $a$ taken
as the `large' parameter, should be substituted to the detailed
enumeration procedure of this Appendix.
Common experience with asymptotic expansions suggests
that  results precise enough for the statistical purposes
of Section 5 should  become available even for fairly low  $a$ values.



\newpage
\begin{figure}
\centering\includegraphics[height=9cm,clip=true]{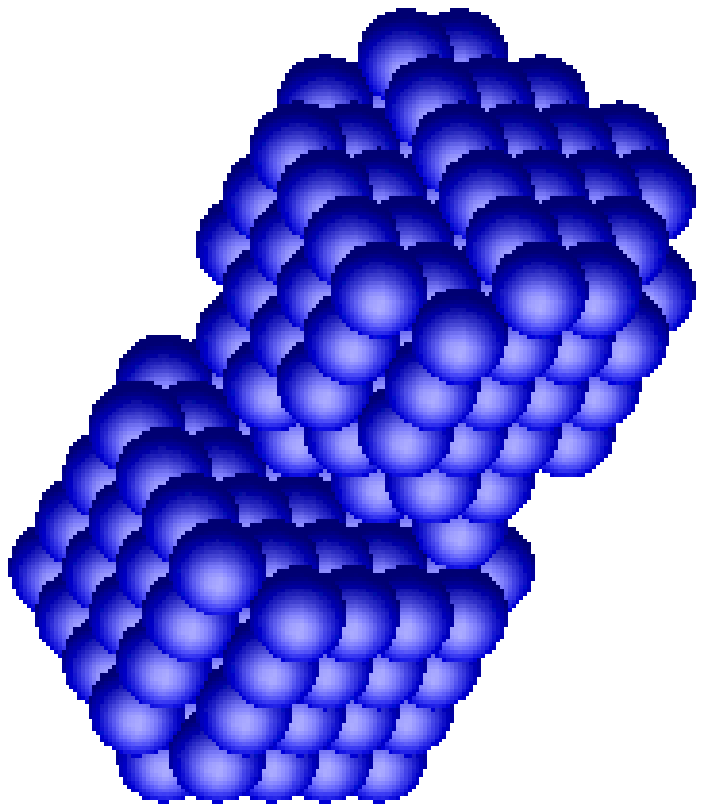}
\caption{ Initial spatial configuration ($t$=0)
for head--on collision of two identical
nuclei, $T$, $I$, of mass 150. Direction of propagation:
$x$--axis. $T$ and $I$ separated by empty layer of cells $x$ = 0.}
\end{figure}

\begin{figure}
\subfigure[]
{\centering\includegraphics[height=7cm,clip=true]{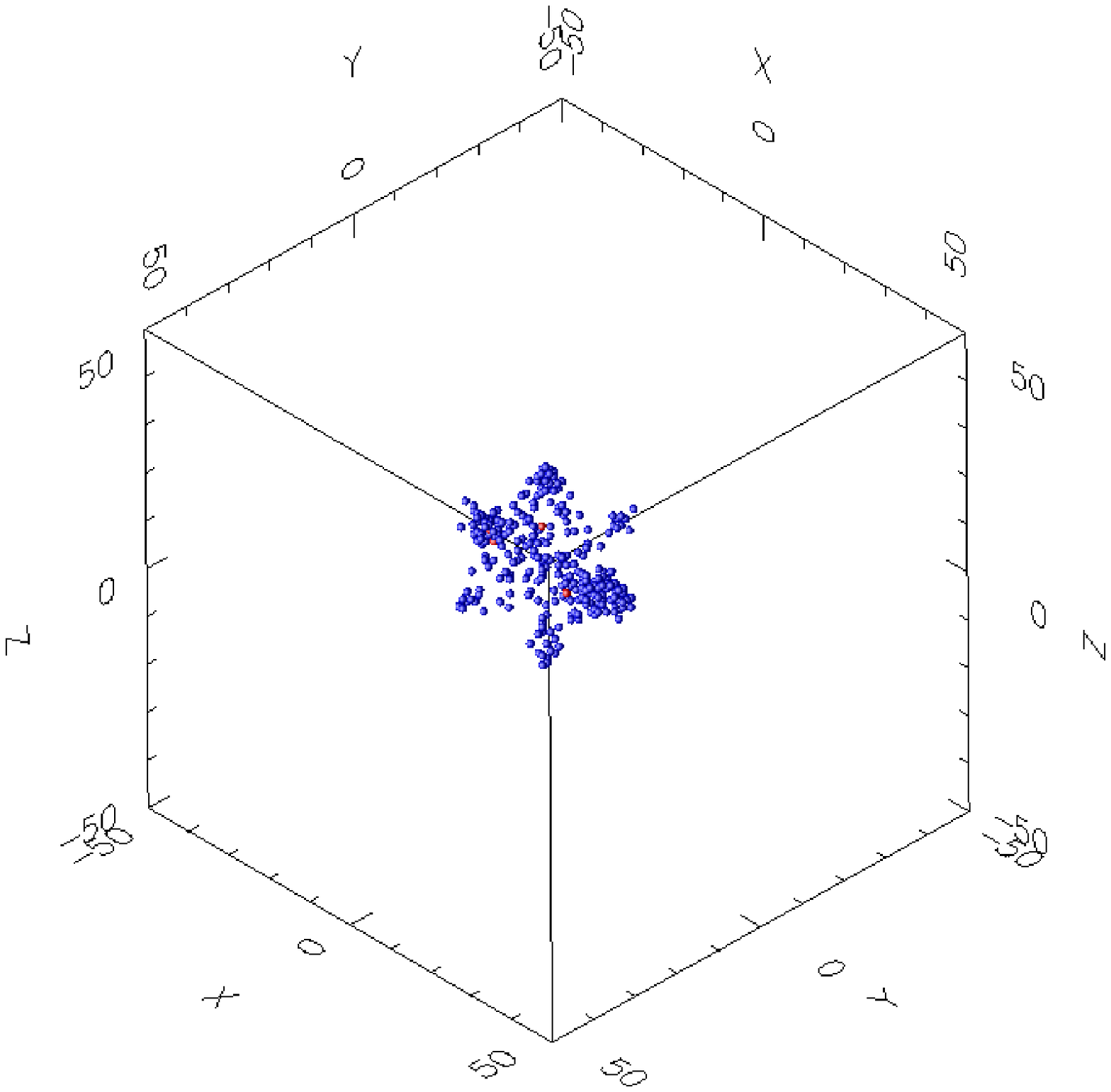}}
\subfigure[]
{\centering\includegraphics[height=7cm,clip=true]{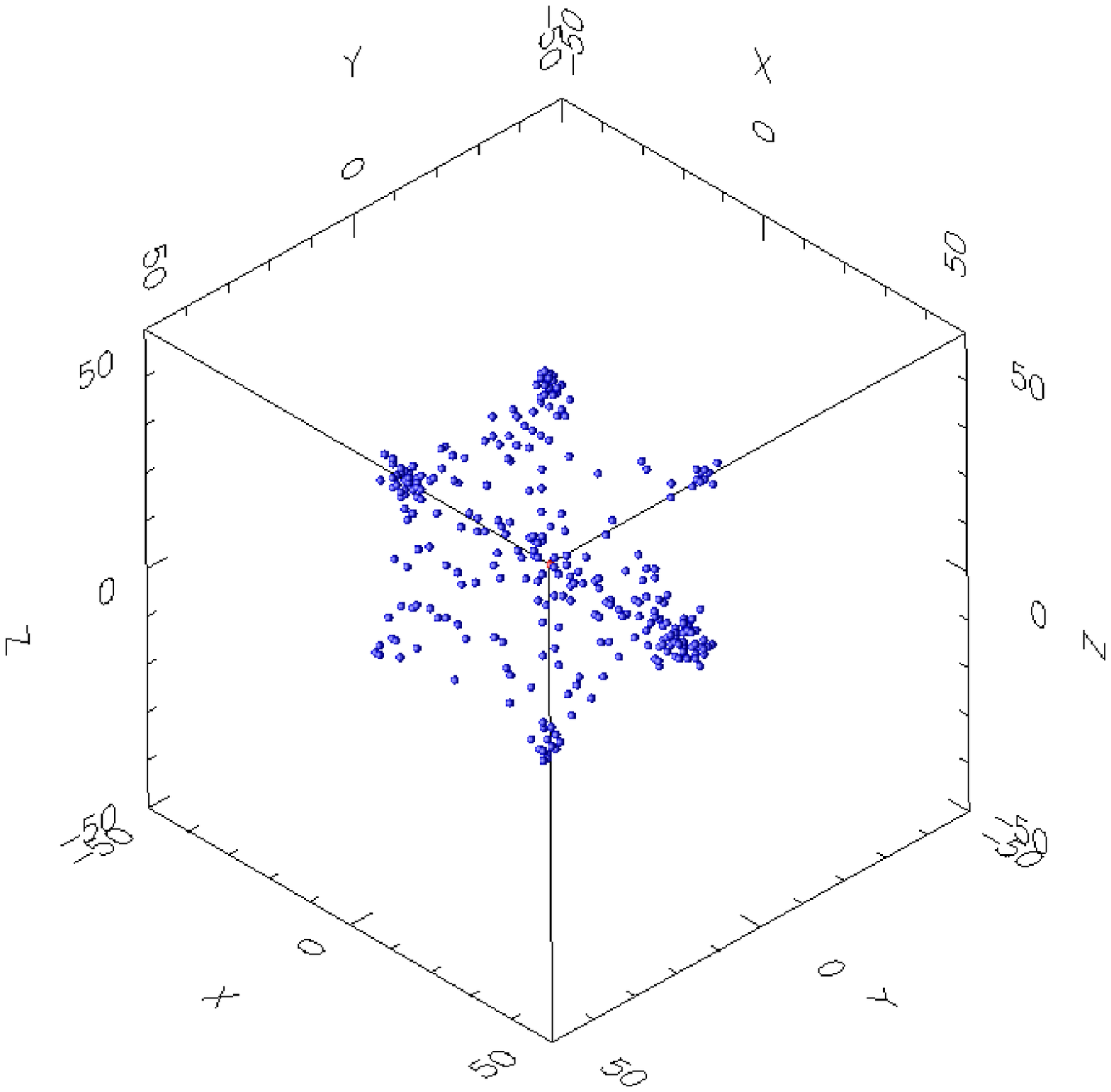}}
\subfigure[]
{\centering\includegraphics[height=7cm,clip=true]{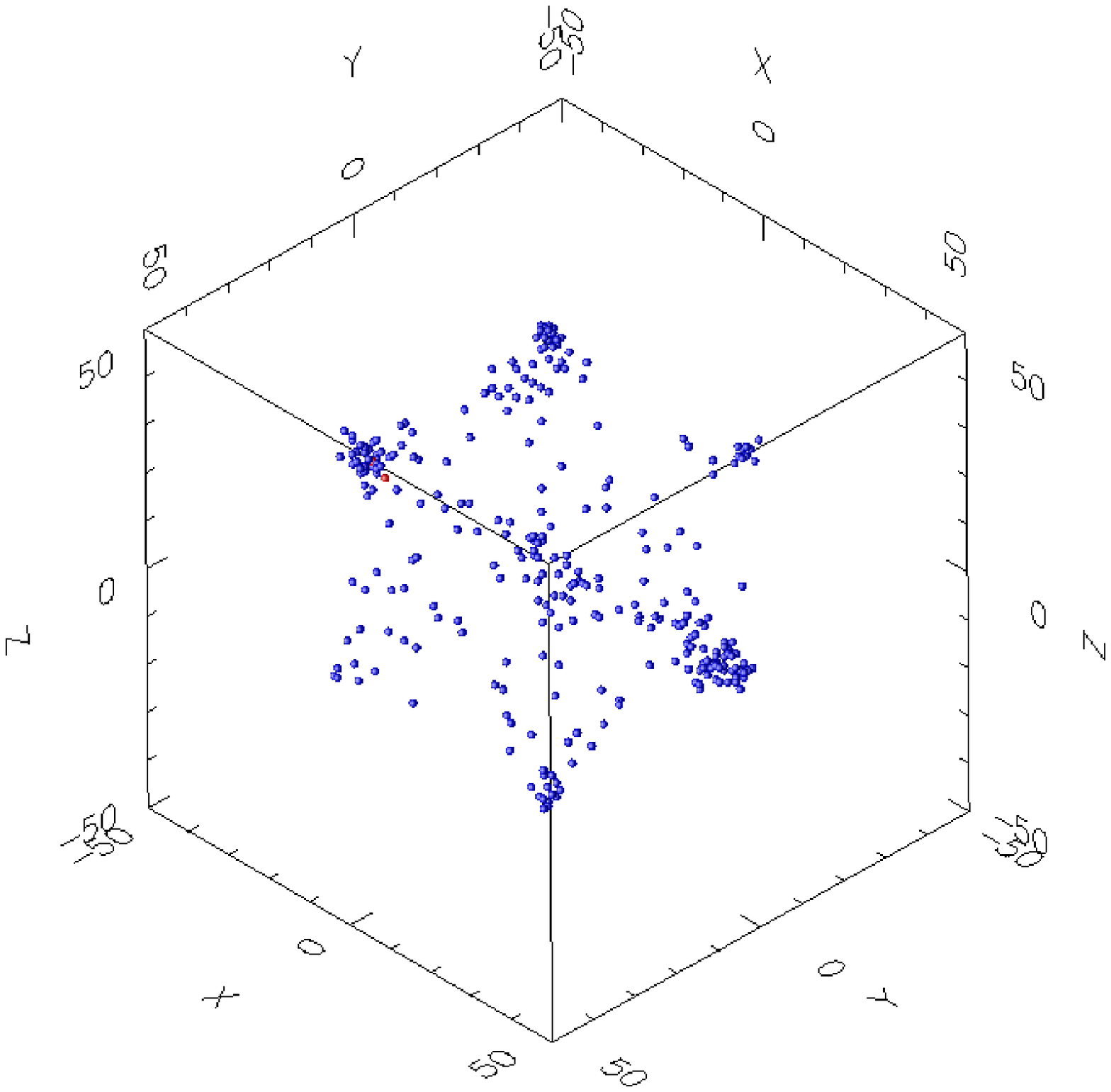}}
\subfigure[]
{\centering\includegraphics[height=7cm,clip=true]{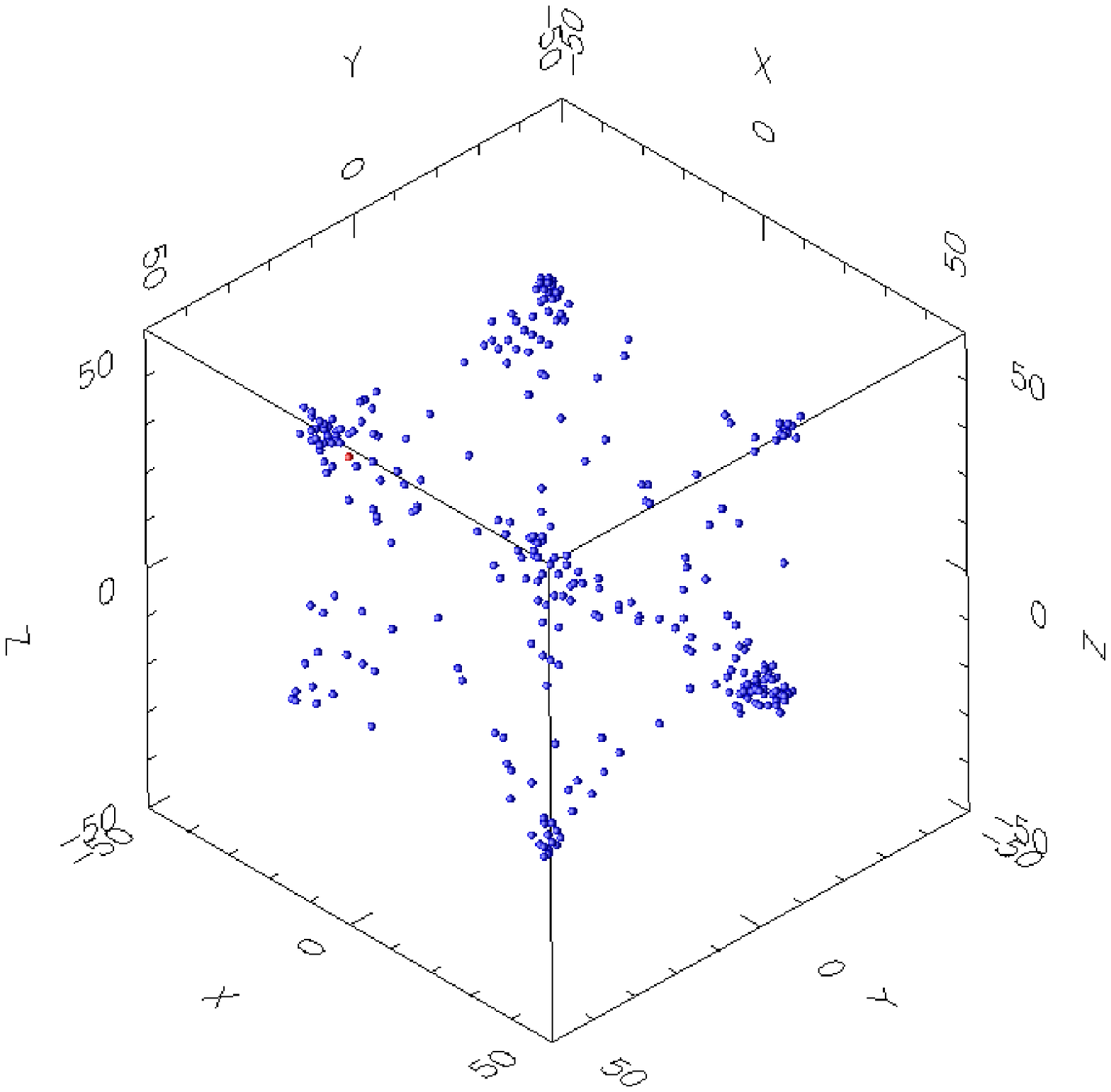}}
\renewcommand{\thesubfigure}{(b')}
\subfigure[]
{\centering\includegraphics[height=4cm,clip=true]{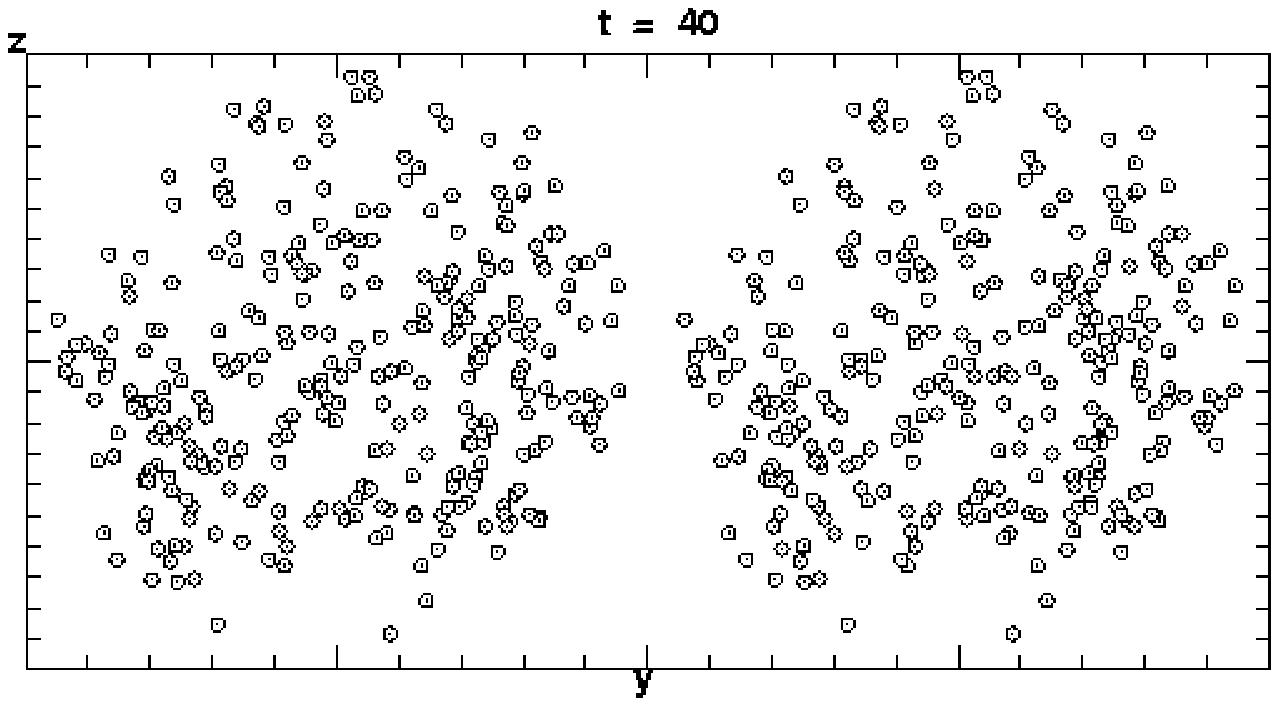}}
\caption{ Configuration of fragments
at timesteps $t$ = 20 (panel a), 40 (b), 50 (c),
and 60 (d). Reconstructed symmetrized distribution of
nucleons for timestep 40 (b$'$).}
\end{figure}

\begin{figure}
\centering\includegraphics[height=12cm,clip=true]{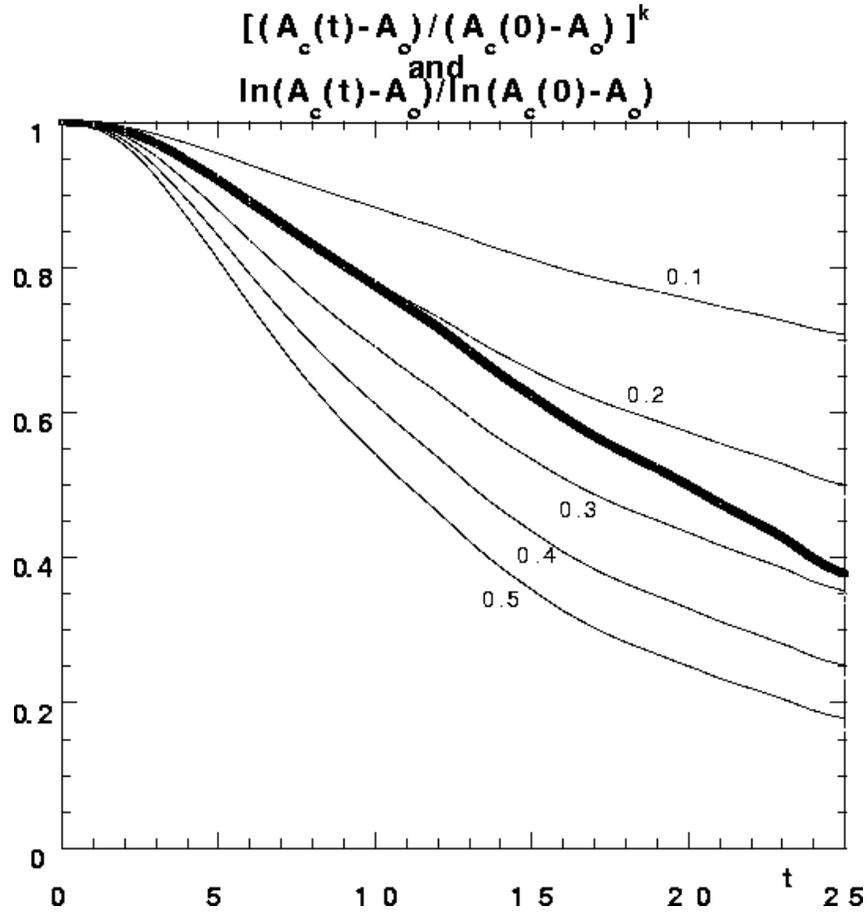}
\caption{ Time--dependence of size of central cluster: $[(A_{C}(t) -
A_{o})/(A_{C}(0) - A_{o})]^{k}$, $k$ = 0.5, 0.4, 0.3, 0.2, 0.1 (thin
curves), and $(\ln(A_{C}(t)-A_{o})/(\ln(A_{C}(0)-A_{o})$ against time
(heavy curve).  Impact energy: 3.75 MeV per nucleon).}
\end{figure}

\begin{figure}
\subfigure[]
{\centering\includegraphics[height=11cm,clip=true]{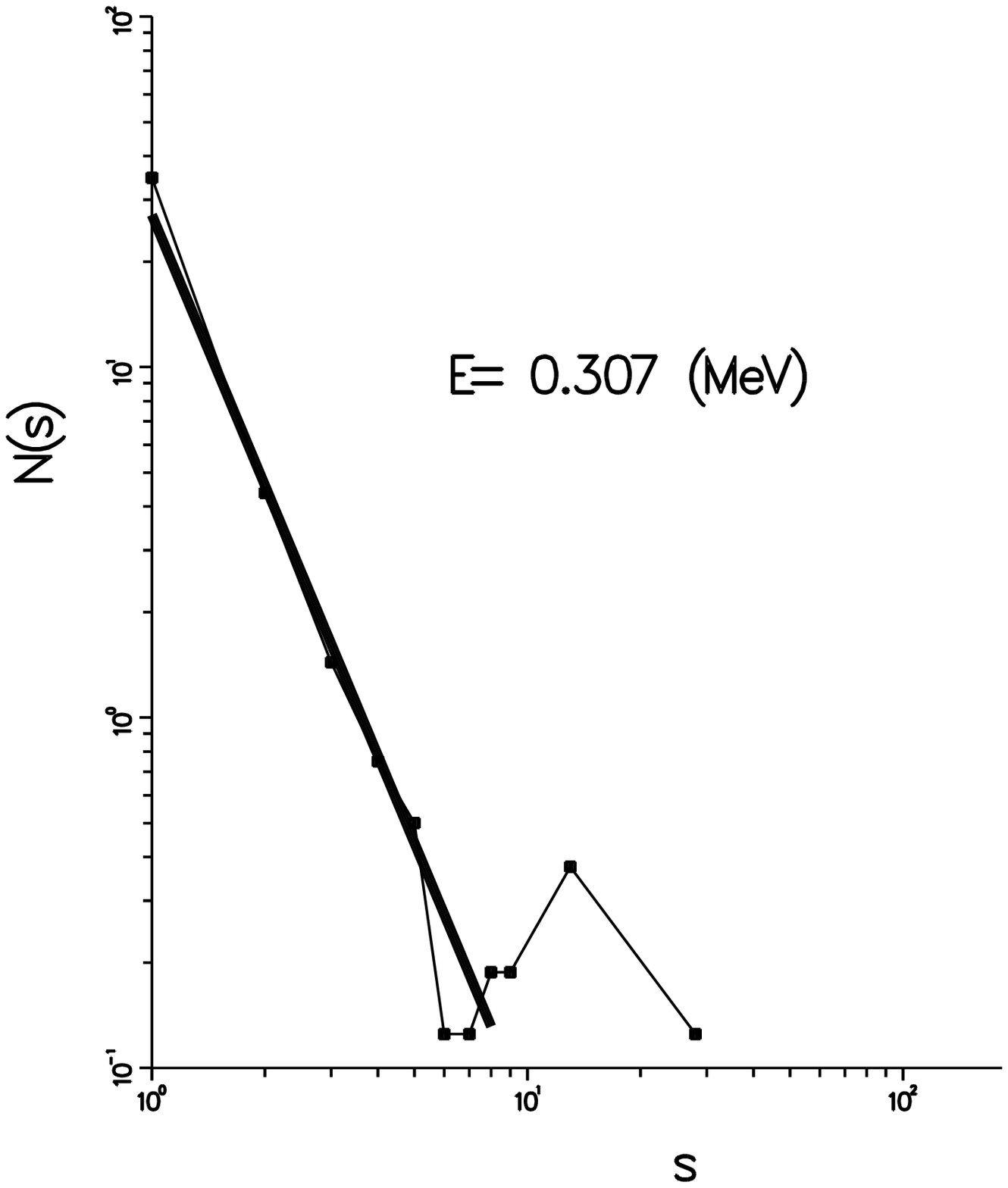}}
\subfigure[]{
\centering\includegraphics[height=11cm,clip=true]{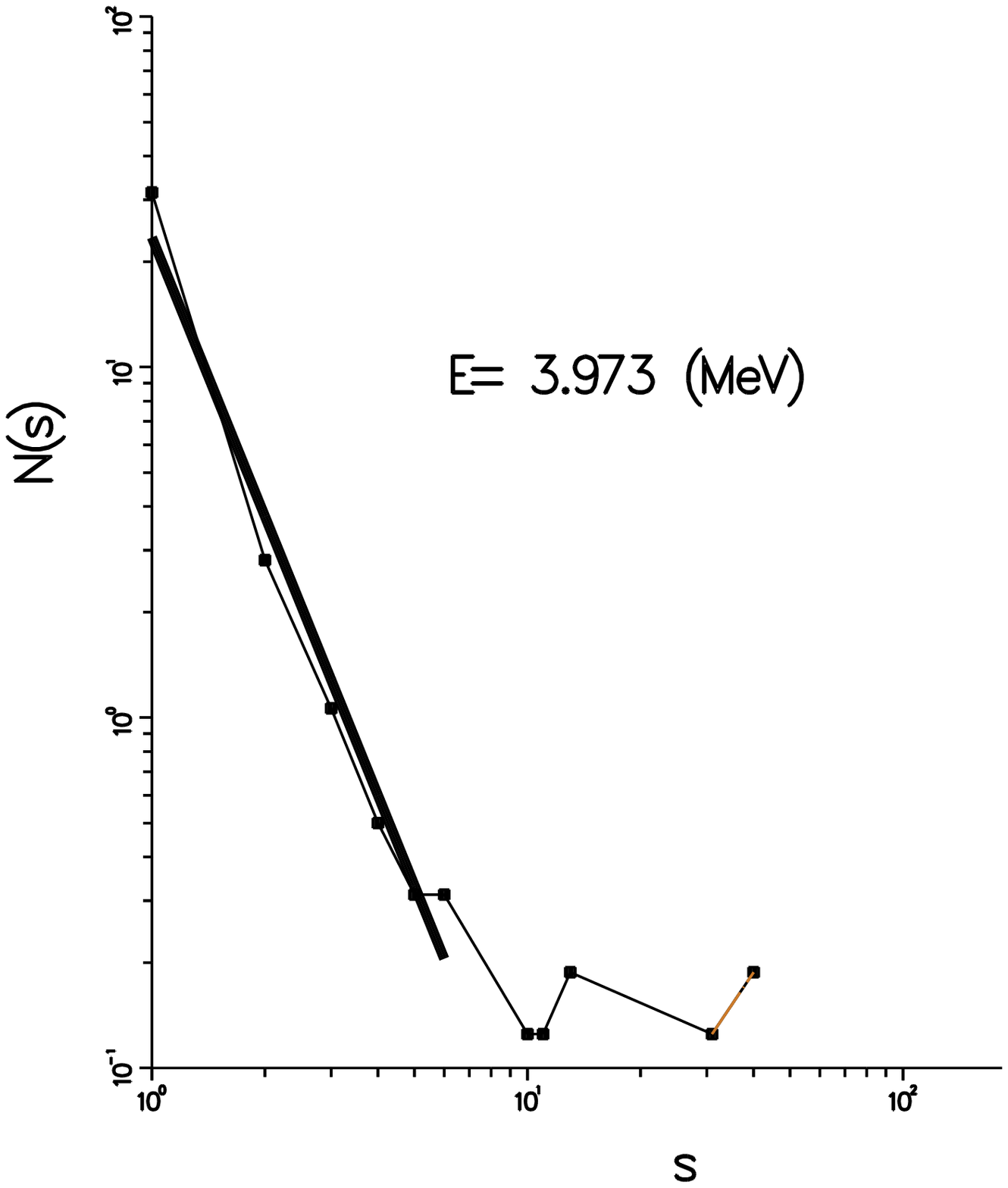}}
\caption{ Cluster distribution $\log$ $N$ -- $\log$ $a$
derived from simulated $4\pi$ detectors,
(a) $E_{imp}/A$ = 0.307 MeV per nucleon: $\tau$ = 2.56;
(b) $E_{imp}/A$ = 3.973 MeV per nucleon: $\tau$ = 2.64.
The numbers of clusters are normalized to the numbers of runs. }
\end{figure}

\begin{figure}
\centering\includegraphics[height=15cm,clip=true]{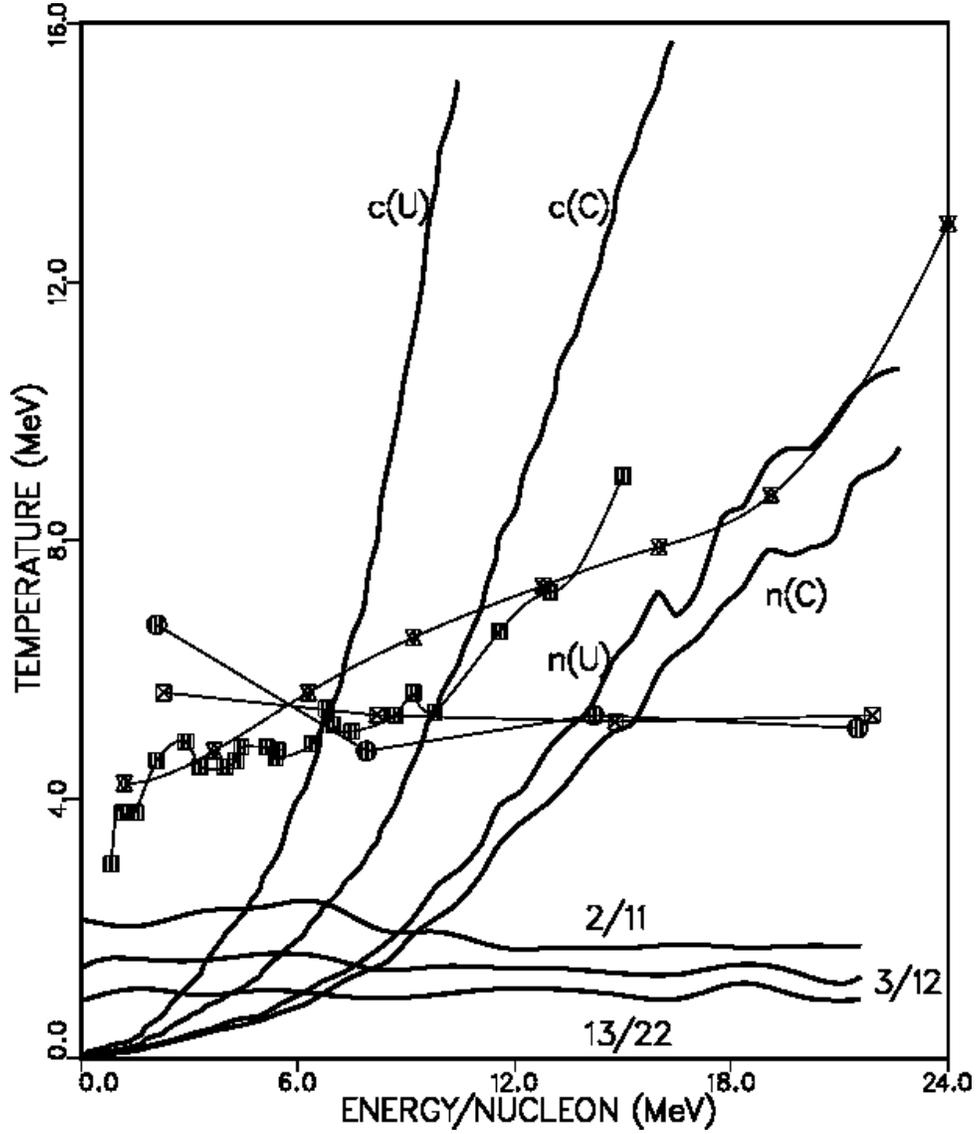}
\caption{ Formal CA caloric curves, $T$--$E_{imp}/A$,
for different definitions of temperature:
(a) $T_{n(C)}$, nucleon--gas in compound nucleus; ($a'$) $T_{n(U)}$,
nucleon--gas in CA universe; (b) $T_{c(C)}$, cluster--gas in compound
nucleus; ($b'$) $T_{c(U)}$;cluster--gas in CA universe; (c)
$T_{13/22}$, reaction equilibrium $ C(1) + C(3) = 2 C(2)$;
($c'$) $T_{3/12}$, reaction equilibrium $ C(3) = C(1) + C(2)$;
($c''$) $T_{2/11}$, reaction equilibrium $ C(2) = 2 C(1)$.
Individual points: experimental estimates from
Au--Au fragments (d) NuPECC interpretation of data (full squares); 
Trautmann interpretation (e, full dots), (f, crossed squares), 
(g, double crosses).}
\end{figure}


\end{document}